\documentclass[11pt,oneside,letterpaper]{article}
\usepackage{eurosym}
\usepackage{graphicx,amsmath,amsfonts}
\usepackage{setspace}
\usepackage{amssymb}
\usepackage{dsfont}
\usepackage{amsmath}
\usepackage{amsfonts}
\usepackage{ifthen}
\usepackage{mathtools}
\usepackage{graphicx}
\usepackage{enumerate}
\usepackage{enumitem}
\usepackage{color}
\usepackage{framed}
\usepackage[margin=1in]{geometry}
\usepackage{natbib}
\usepackage[dvipsnames]{xcolor}
\usepackage[colorlinks=true, urlcolor=Mahogany, linkcolor=Mahogany, citecolor=Mahogany]{hyperref}
\usepackage{fp}
\usepackage[abspage,user,savepos]{zref}
\usepackage[capitalize]{cleveref}
\usepackage{amsthm}
\usepackage{caption}
\usepackage{subcaption}
\usepackage{palatino}
\usepackage{cuted,balance}

\usepackage[ruled]{algorithm2e}

\usepackage{csquotes}

\makeatletter
\renewenvironment*{displayquote}
  {\begingroup\setlength{\leftmargini}{0.5cm}\csq@getcargs{\csq@bdquote{}{}}}
  {\csq@edquote\endgroup}
\makeatother

\newtheorem{definition}{Definition}
\newtheorem{proposition}{Proposition}
\newtheorem{theorem}{Theorem}
\newtheorem{lemma}{Lemma}
\newtheorem{remark}{Remark}

\usepackage[ruled]{algorithm2e}

\newcommand{\buyerset}{\mathcal{N}}
\newcommand{\vbf}{\mathbf{v}}
\newcommand{\buyerCDF}{G}
\newcommand{\buyerPDF}{g}
\newcommand{\virtualV}{\phi_{\buyerCDF}}
\newcommand{\ironvirtualV}{\bar{\phi}_{\buyerCDF}}
\newcommand{\levelNum}{k}
\newcommand{\pbf}{\mathbf{p}}
\newcommand{\bidset}{\mathcal{P}}
\newcommand{\eqbid}{b^*}
\newcommand{\rev}[2]{\textsc{Rev-ALG}_{\buyerCDF}\left(#1,#2\right)}
\newcommand{\opt}[1]{\textsc{Rev-OPT}_{\buyerCDF}\left(#1\right)}
\newcommand{\PriceThresh}{\Gamma_{\buyerCDF}}
\newcommand{\ThreshPrice}{\PriceThresh^{-1}}
\newcommand{\maxV}{\textrm{OPT}}
\newcommand{\algT}{\textrm{ALG}}

\let\originaleqref\eqref
\renewcommand{\eqref}{\originaleqref}

\begin{document}

\title{Descending Price Auctions with Bounded Number of Price Levels and Batched Prophet Inequality}

\author{Saeed Alaei  \thanks{%
Google Research. \texttt{saeed.a@gmail.com}} \and Ali Makhdoumi \thanks{%
Fuqua School of Business, Duke University. \texttt{ali.makhdoumi@duke.edu}}
\and Azarakhsh Malekian \thanks{%
Rotman School of Management, University of Toronto. \texttt{%
azarakhsh.malekian@rotman.utoronto.ca}} \and Rad Niazadeh \thanks{%
The University of Chicago Booth School of Business \texttt{rad.niazadeh@chicagobooth.edu}}}

\maketitle

\begin{abstract}

 
We consider descending price auctions for selling $m$ units of a good to unit demand i.i.d. buyers where there is an exogenous bound of $k$ on the number of price levels the auction clock can take. The auctioneer's problem is to choose price levels $p_1 > p_2 > \cdots > p_{\levelNum}$ for the auction clock such that auction expected revenue is maximized. The prices levels are announced prior to the auction. We reduce this problem to a new variant of prophet inequality, which we call \emph{batched prophet inequality}, where a decision-maker chooses $k$ (decreasing) thresholds and then sequentially collects rewards (up to $m$) that are above the thresholds with ties broken uniformly at random. For the special case of $m=1$ (i.e., selling a single item), we show that the resulting descending auction with $k$ price levels achieves $1- 1/e^k$ of the unrestricted (without the bound of $k$) optimal revenue. That means a descending auction with just 4 price levels can achieve more than 98\% of the optimal revenue. We then extend our results for $m>1$ and provide a closed-form bound on the competitive ratio of our auction as a function of the number of units $m$ and the number of price levels $k$.


\end{abstract}

\section{Introduction}
\label{sec:intro}
\newcommand{\DCA}{descending price auction }
\newcommand{\dca}{DPA}

The \DCA{}(\dca{}) is strategically equivalent to the celebrated first price auction, a powerful machinery in auction design that has passed the test of time in the history of practical auctions~\citep{milgrom2004putting}. In a stark contrast to posted pricing mechanisms, this auction induces competition among the buyers. This competition in turn helps with increasing the generated revenue --- to the extent that in a symmetric independent private value setting for selling a single item this auction (with an appropriate reserve price) is revenue optimal. While there are other auction formats, such as the second price auction, that induce competition and enjoy similar revenue benefits compared to \dca{}, there is a long list of reasons why \dca{} is more practically appealing:\footnote{In a recent trend, Google has reported switching their Ad Manager (for advertising exchange), AdMob (for mobile advertising), and more recently AdSense (for website monetization), all of which are cornerstones of Google’s publisher-facing toolset for display ads, to variants of the first price auction.} pay-as-bid is a simple, transparent, and interpretable rule, and makes the auctioneer more credible as descending price auctions are ``automatically self-policing''~\cite{vickrey1961counterspeculation,akbarpour2020credible}. Furthermore, buyers reveal the minimum information about their private valuations in \dca{} compared to other auction formats. The major disadvantage of \dca{}, as with other clock auctions, is its requirement of many rounds of communication between the auctioneer and the buyers.

Another simple and highly prevalent selling mechanism in practice is posted pricing. Notably, posted pricing requires the minimum communication between the auctioneer and the buyers and reveals even less information about the buyers valuations than \dca, and is even a simpler mechanism in many ways. At the same time, it can be perceived as a special case of \dca{} with a single price level. As a result, it enjoys almost all other practical appeals of \dca{} such as transparency and credibility; however, its performance is not optimal (both in terms of revenue and welfare) due to the lack of competition among the buyers.

The above paradigm motivates us to study the impact of having a constraint on the number of rounds of \dca{} and to evaluate how this constraint changes the competition, and hence the revenue/welfare performance of the resulting mechanism. 

To this end, we consider the class of descending price auctions with a \emph{bounded number of price levels}. In this new auction format, the auctioneer commits to a distinct set of $k\in\mathbb{N}$ prices $p_1>p_2>\ldots>p_k$ for an exogenously specified $k$. During each round $i\in[k]$ of the auction, the remaining supply is offered for sale at price $p_i$ to all interested buyers in a random order. All sales are private and not revealed to other buyers. The auction terminates once all units are sold or at the end of round $k$ whichever happens first. This auction interpolates between the \DCA{} with a reserve price, which can achieve the optimal revenue, and anonymous posted pricing, which is sub-optimal in general. We then ask the following design questions:

\begin{displayquote}
\emph{By using at most $k$ distinct price levels, how well can a $k$-level \DCA{} approximate the optimal mechanism? Is there a simple sequence of $k$ prices that can approximate the optimal mechanism in a parametric fashion?}
\end{displayquote}

In this paper, we answer these questions in affirmative. In particular, we consider a general setting with multiple unit-demand symmetric buyers with independent private values and multiple identical items. We then show how to design a simple sequence of prices to approximate the optimal mechanism. We also quantify the extent of the approximation as a function of the number of  price levels. Interestingly, we show that the revenue of a $k$-level \dca{} with properly designed prices converges to the optimal revenue exponentially fast as $k$ goes to infinity. This result greatly extends the applicability of descending price auctions with bounded number of price levels, as the optimal mechanism is very well approximated by using this class of mechanisms, even for small values of $k$. We also obtain similar results for the welfare objective through a different sequence of prices.

\subsection{Main Contribution}
We start by formulating the descending price auction with $k$ price levels. In this problem, there are $n \ge 1$ unit-demand buyers whose values are drawn i.i.d. from a known distribution, and an auctioneer who sells $m$ units of a good. The auctioneer announces a sequence of $k$ (decreasing) prices and then, after observing their values, buyers decide about their bid (from this set of prices). The auctioneer then goes over the prices in a decreasing order over $k$ rounds. If the bid of multiple buyers are equal to the posted price in a round, the (remaining) items go to a subset of them chosen uniformly at random. Finally, the auctioneer charges each winning buyer with her submitted bid.

After establishing the existence of a (symmetric) Bayes-Nash equilibrium for buyers and characterizing its properties, we turn to our main question: what is the revenue approximation factor of a descending price auction with $k$ posted prices in selling $m\in\mathbb{N}$ items. The mathematical definition of approximation in this paper is a standard one from the design and analysis of algorithms and mechanisms (see, e.g., \cite{hartline2012approximation}). For $\Gamma\in[0,1]$, a $k$-\dca~ is a $\Gamma$-approximation for the setting with $m$ items and multiple buyers with i.i.d. values if for all instances of this problem the performance, i.e., the expected revenue, of $k$-\dca~ is within a
multiplicative $\Gamma$ fraction of the performance of the optimal mechanism for that instance.

To guide the analysis, we introduce a new class of prophet inequality~\citep{krengel1978semiamarts,samuel1984comparison}, and refer to it as the \emph{batched prophet inequality}. In this new problem, all random rewards are drawn at time $0$. The decision-maker decides about $k$ different thresholds for $k$ rounds and can take $m$ rewards. At each round if all the
rewards are below that round's threshold, the game advances to the next round. Otherwise, the decision-maker wins
all of the rewards that has passed the threshold. If the number of the rewards above the threshold is larger than the remaining number of rewards (out of $m$) that the decision-maker can take, then she only collect a subset of these rewards to fill her capacity $m$, breaking ties in this selection uniformly at random. The game proceeds until the decision-maker selects $m$ rewards or after $k$ rounds. The goal of the decision-maker is to choose $k$ thresholds such that the expected value of the collected rewards is a good approximation of the expected value of the sum of top $m$ rewards (i.e., the expected reward when all the variables are known).

Our first main result, stated next, proves a reduction form the revenue approximation problem for $k$-\dca~ to the batched prophet inequality problem.

\medskip
\noindent \textbf{Main Result 1 (informal):} \emph{Any algorithm for the batched prophet inequality with $k$ rounds and capacity $m$ that achieves $\Gamma(k,m)$ approximation of the expected sum of top $m$ rewards can be used to generate a sequence of $k$ prices for the $k$-\dca~ whose expected revenue is $\Gamma(k,m)$ approximation of the optimal expected revenue for selling $m$ items. }
\medskip

Building on this result, we then develop the analysis of the batched prophet inequality for $m=1$ item and $m>1$ items separately as they require different techniques. 
The following summarizes our result for a single item.

\medskip
\noindent \textbf{Main Result 2 (informal):}\emph{ There exists a sequence of $k$ thresholds that achieves $1-1/e^k$ of the optimal expected reward for the batched prophet inequality. This implies the existence of $k$ prices, so that their corresponding $k$-\dca~ achieves $1- 1/e^k$ of the optimal revenue. }

\medskip

The main idea of the proof is to  construct a sequence of prices that balances the tradeoff between having a high revenue from the current round (by selecting a large threshold in the batched prophet problem) and having a high probability for collecting future rewards (by selecting a small threshold in the batched  prophet problem). There are two points worth mentioning. First, we provide an algorithm that explicitly finds the sequence of thresholds that achieves the competitive ratio $1- 1/e^k$ for the batched prophet inequality. Second, we also explicitly characterize the sequence of prices for $k$-\dca~ that achieves $1- 1/e^k$ of the optimal revenue. Third, our approximation factor for the batched prophet inequality is  optimal for $k=1$. In particular, for $k=1$, our batched prophet inequality becomes identical to the classic prophet inequality for i.i.d. rewards (or the homogeneous prophet secretary problem) with a static policy: the sequence of random variables arrive at a random order and the decision-maker selects the first reward that is above a static threshold. This problem has been studied in \cite{correa2017posted}, \cite{ehsani2018prophet}, and \cite{lee2018optimal}, where the authors establish the existence of a distribution for rewards such that no single threshold policy can achieve better than $1- 1/e$ of the optimal expected reward.\footnote{These papers also show there exists simple static thresholds achieving this ratio.}

We then consider the generalization of our setting to selling multiple items (i.e., $m>1$). Again, we first prove that the problem of approximating the optimal revenue reduces to a batched prophet inequality in which the decision-maker can acquire up to $m$ rewards. We then provide an algorithm for approximating the optimal expected reward in the batched prophet inequality. The following summarizes our result for multiple items.

\medskip
\noindent \textbf{Main Result 3 (informal):}\emph{ For any $\epsilon>0$, there exists a sequence of $k$ thresholds that achieves 
\begin{align*}
     \left(1- e^{-m} \frac{m^m}{m!} \right) + \sum_{r=1}^{k-1} \sum_{i=0}^{m-1} \frac{m^{i} r^{i}}{i!} e^{-mr} \left(1- e^{-(m-i)} \frac{(m-i)^{m-i}}{(m-i)!} \right) - \epsilon
\end{align*}
of the optimal expected reward for the batched prophet inequality when the number of awards (corresponding to buyers) is large enough. This implies the existence of $k$ prices, so that their corresponding $k$-\dca{} achieves the same approximation of the optimal revenue.}
\medskip

Again, a noteworthy point is that our approximation factor for the batched prophet inequality with $m>1$ items is also optimal for $k=1$. In particular, for $k=1$ and $m>1$, our batched prophet inequality problem becomes identical to the prophet inequality for i.i.d. rewards (or the homogeneous prophet secretary problem) with a static policy when the decision-maker can collect $m$ rewards. This problem has been studied in \cite{yan2011mechanism} and \cite{arnosti2021tight}, where the authors establish the existence of a distribution for rewards such that no single threshold policy can achieve better than $1- e^{-m} \frac{m^m}{m!}$ of the optimal expected reward, and show this approximation factor can be achieved by a single threshold. As a side note, our analysis for the setting with multiple items can be viewed as an extension of the results of \cite{arnosti2021tight} to any $k>1$ by using a different, and arguably simpler, analysis.

Our main focus is on approximating the optimal revenue. However, we show that our analysis provides the above approximations for the optimal welfare as well.

\subsection{Further Related Literature}
Besides the papers discussed earlier in the introduction, our paper is related to the vast literature on Bayesian mechanism design in computer science, economics, and operations. We sketch some of these connections below.
\smallskip
\paragraph{Simple vs. optimal} Our paper relates to the literature that studies simple mechanisms that approximates the optimal mechanism; Representative papers are \cite{chawla2007algorithmic}, \cite{hartline2009simple}, \cite{chawla2009sequential}, \cite{cai2011extreme}, \cite{yan2011mechanism},  \cite{haghpanah2015reverse},  \cite{alaei2019optimal}, and \cite{jin2019tight} that establish the approximation factor of this mechanism and study other variants of (sequential) posted pricing under a variety of assumptions on the underlying value distributions. In particular, \cite{chawla2007algorithmic} consider a unit-demand multi-dimensional screening problem and identify a simple item pricing that is approximately optimal and \cite{alaei2019optimal} proves that an anonymous posted price can achieve $1/e$ of the optimal auction for regular, independent, and non-identical values (see also \cite{hartline2012approximation} for a survey of earlier works in this literature). We depart from this literature by stepping away from posted pricing and studying the performance of a descending clock auction with multiple price levels. As we establish in the paper, both our analysis and results are different from the above papers.
\smallskip
\paragraph{First price/descending clock auctions with discrete bids/price levels} On one extreme of our auction formats, we have anonymous posted pricing whose (sub)optimality has been extensively studied in the literature, as mentioned earlier. On the other extreme, and more closely to ours, are \cite{chwe1989discrete} and  \cite{horner2011managing} that consider first price auction with discrete bids and prove that when the distribution of the underlying values is uniform and the discrete bids (over time) are multiples of an increment, as the number of discrete bids goes to infinity the auctioneer's revenue converges to the optimal revenue.
Our paper considers the interpolation between these two extremes and aims to understand the performance guarantee of a \DCA  with multiple posted prices. In particular, we depart from these papers by asking the following question: what is the ``best'' approximation that the auctioneer can achieve by considering $k$ discrete bids (that are not restricted to be multiples of an increment)? By building connections to the batched prophet inequality problem, we establish that, for general distributions, by properly choosing the sequence of discrete bids the auctioneer's revenue converges exponentially fast to the optimal revenue and we characterize its parametric convergence rate in terms of the number of price levels and items.

Another work that is related to ours in spirit is \cite{nguyen2014optimizing}. They consider a framework for optimizing prices in a multi-item descending clock auction, when auctioneer is a buyer and selecting sellers providing different items within a feasibility constraint, in order to minimize expected payment. Our work diverges from this paper as we consider a different setting and objective (optimizing prices under feasibility vs. approximation ratio analysis with respect to optimal auction); however, the \emph{percentile-based} price decrements has the same flavor as our price trajectories in \Cref{sec:final-price}.

\smallskip
\paragraph{Prophet inequality} Our paper also relates to the rich literature on prophet inequality. In the vanilla version of this problem, a decision-maker sequentially observes rewards drawn independently from known distributions, and decides when to stop and take the reward to maximize her expected collected reward. 
Prophet inequality was first introduced and analyzed in \cite{krengel1978semiamarts}, \cite{hill1982comparisons}, and \cite{samuel1984comparison} 
and further developed in \cite{babaioff2007matroids}, \cite{kleinberg2012matroid}, \cite{azar2014prophet},
\cite{dutting2020prophet}, and \cite{liu2020variable} among others. We depart from this literature by establishing that our $k$-level \dca~ reduces to a batched version of prophet inequality, as described earlier. 
This problem differs from the prophet inequality in two main ways: (i) unlike prophet inequality where random variables are realized and observed by the decision-maker sequentially over time, here all random variables are realized at time $0$ and the decision-maker does not observe the value of any of the rewards throughout the game. (iii) unlike the prophet inequality that does not have recall (i.e., the decision-maker cannot acquire a variable that was realized in previous rounds), all random variables are realized in a batch time $0$ and the decision-maker may (randomly) acquire a variable at any round if its value is the above the threshold and all random variables (including this one) were below the posted thresholds in previous rounds.

Besides the classic prophet inequalities, many variations
such as prophet inequalities with limited samples form the distributions \cite{azar2014prophet}, static policies for prophet secretary \cite{correa2017posted, chawla2020static, arnosti2021tight} (also see \cite{correa2021prophet} that introduce the blind strategies, which is a generalization of the static policy to a multi-threshold setting), i.i.d. and random order \cite{abolhassani2017beating,esfandiari2017prophet}, and ordered prophets (a.k.a. the free order sequential posted pricing problem) \cite{azar2018prophet, beyhaghi2018improved, correa2021prophet} have been explored. Also, 
establishing connections to the price of anarchy \cite{dutting2020prophet} and online contention resolution
schemes \cite{feldman2016online} have been of particular interest in this literature. Generalizations of the simple prophet inequality
problem to combinatorial settings have also been studied, where the examples are matroids \cite{krengel1978semiamarts, ehsani2018prophet} and knapsack \cite{feldman2016online, jiang2022tight}. For a survey of recent developments in prophet inequality see \cite{correa2019recent} and for connections to mechanism design see \cite{lucier2017economic}.

\medskip
\noindent\textbf{Organization:} The rest of the paper proceeds as follows. In Section \ref{sec:model}, we present our problem formulation and formally define \DCA with $k$ price levels (which we call $\levelNum$-\dca{}). We then introduce batched prophet inequality and prove that $\levelNum$-\dca{} reduces to it. In Section \ref{sec:single-unit}, we consider batched prophet inequality with a single item and prove that the decision-maker can achieve $1- 1/e^k$ of the optimal expected reward. We then use this solution to design a $\levelNum$-\dca{} with the same approximation factor for the optimal expected revenue. In Section \ref{sec:multiple}, we extend our analysis of the batched prophet inequality and $\levelNum$-\dca{} to a setting with multiple items. Section~\ref{sec:conclusion} concludes, while the Appendix presents the omitted proofs from the text.









\section{Problem Formulation, Equilibrium, and  Reduction}
\label{sec:model}
We start by formalizing our setting (\cref{sec:model:environment}) and revenue maximization problem (\cref{sec:problem}). We then characterize the equilibrium behaviour of the buyers (\cref{sec:equilibrium}) and show how our problem reduces to a variant of the prophet inequality problem (\cref{sec:reduction}).
\subsection{The Environment}
\label{sec:model:environment}
 We consider a symmetric single-parameter Bayesian mechanism design setting, in which an auctioneer is selling one or multiple units of the same item to $n$ unit-demand buyers --- represented by the set $\buyerset=\{1,2,\ldots,n\}$ --- to maximize the expected revenue. We rely on standard definitions, solution concepts, and results in Myerson's theory on the design of optimal single-parameter auctions~\citep{myerson1981optimal}, e.g., see Chapter~3 of \cite{hartline2013mechanism}. In particular, we assume buyers' utilities are quasi-linear, meaning that given the value $v$ from receiving the item and a payment $p$, the buyer's utility is $v-p$. We assume buyers' private values $\vbf=(v_1,v_2,\ldots,v_n)$ for the item are non-negative and drawn i.i.d. from a known common prior distribution with a commutative distribution function (CDF) $\buyerCDF:\mathbb{R}_{\geq 0}\rightarrow [0,1]$, and this distribution admits a probability density function (PDF) $\buyerPDF:\mathbb{R}_{\geq 0}\rightarrow [0,1]$. For simplicity of technical exposition, we further assume this distribution is atom-less and hence $\buyerCDF$ is continuous.\footnote{The fact that the CDF is continuous has some bearing on the results: without this assumption, and when the distribution has atoms, our guarantees can only remain valid after incorporating a standard randomized tie-breaking trick into our mechanisms or after adding small perturbations to the underlying value distribution.}  Given a buyer's value distribution, the Myerson's virtual value function is defined as $\virtualV(v)=v-\frac{1-\buyerCDF(v)}{\buyerPDF(v)}$. We restrict our attentions to regular distributions for which the virtual value function is weakly-increasing.\footnote{This assumption is without loss of generality, as we can replace $\virtualV$ with the Myerson's ironed virtual valuation function $\ironvirtualV$ for irregular distributions, which is weakly-increasing~\citep{myerson1981optimal}. As Myerson's payment equivalence lemma holds for ironed virtual values, all of our results remain intact. We omit the details for brevity.}
 
 The auctioneer is interested in running a \emph{ $\levelNum$-level descending price auction ($\levelNum$-\dca{})} to maximize her expected revenue. This auction, formally defined below, is basically a descending price (clock) auction when the price levels, and therefore the bid of buyers, are restricted to a finite set of cardinality $\levelNum$.
 \begin{definition}[$\levelNum$-\dca{}] Given a finite set of distinct price levels (also referred to as bids) $\bidset=\left\{p_1,p_2,\ldots,p_{\levelNum}\right\}$, where $p_1> p_2 > \ldots > p_{\levelNum}$, its corresponding $m$-unit $\levelNum$-\DCA for $m\in\mathbb{N}$ is the following mechanism:
 \begin{itemize}
     \item Ask each buyer $i$ to place a bid $b_i\in\mathcal{P}\cup\{0\}$ after observing her private value $v_i\sim \buyerCDF$,\footnote{Bidding $b_i=0$ is always a possibility to guarantee the individual rationality of our auctions.}
     \item Select the set of winners by greedily picking the top $m$ submitted bids (and breaking the ties uniformly at random),
     \item Charge each buyer $i$ in the winner set with her submitted bid $b_i$.
 \end{itemize}
  \end{definition}
 As a remark, $k$-\dca{} with $k=1$ is basically an anonymous pricing mechanism with a randomized tie-breaking rule. Also, when $k=+\infty$ and $\bidset=[p,+\infty)$, this auction boils down to ordinary descending price auction with an anonymous reserve price of $p$. In this case, by setting $p={\virtualV}^{-1}(0)$, we can recover the Myerson's optimal auction. 
 
 Importantly, an alternative interpretation of the above mechanism is basically a sequential pricing when the buyers are flexible to decide on their purchase round: in such a setting, $n$ i.i.d. buyers arrive at round $0$ and the seller posts a (decreasing) sequence of prices $\pbf=(p_1,\ldots,p_k)$ to sell $m$ items over a finite horizon of $k$ discrete rounds $1,\ldots,\levelNum$. Now each buyer $i$, given her value $v_i$, decides when to purchase the item (if any). The game starts at round $1$. At each round $j\in[\levelNum]$, if all the units of the item have not been sold so far, the buyers who have decided to purchase at price $p_j$ have a chance to receive one unit of the item at this price (when the ties are broken uniformly at random if there is more than one buyer interested in the last unit of the item). If there is still some unsold units of the item at the end of this round, the game advances to the next round $j+1$. In this alternative interpretation of our setting, the buyers only interact with the auctioneer (in at most $k$ rounds) and do not observe the number of sold items. This assumption is particularly relevant for settings in which the buyers and the auctioneer privately/remotely interact with each other. Throughout the paper, we use both $\pbf=(p_1,\ldots,p_k)$ and $\bidset=\left\{p_1,p_2,\ldots,p_{\levelNum}\right\}$ to denote the prices of $k$-\dca{}.

 \subsection{Auctioneer's Problem}
 \label{sec:problem}
 The auctioneer's objective is to maximize her expected revenue by choosing the price sequence $p_1>\ldots>p_k$, taking into account the buyers' strategies at equilibrium. In particular, once  prices are fixed, the buyers play a game of incomplete information, in which each buyer's strategy is a bid function that maps her private value to one of the possible prices (or a distribution over the possible prices for mixed strategies). We evaluate the expected revenue of the resulting $k$-\dca{} under a \emph{Bayes-Nash equilibrium (BNE)} of this game.  The equilibrium selection (and uniqueness) will be detailed later in \Cref{sec:equilibrium}.
 
 Given the realized values $\vbf$, by abuse of notation, we denote by $$\rev{\mathbf{v}}{\pbf}$$ the revenue of the $k$-\dca{} with prices $\pbf$ at a particular BNE when buyers' values are drawn i.i.d. from a common prior distribution $\buyerCDF$. We also denote by $$\opt{\vbf}$$ the revenue of the optimal direct mechanism when buyers' values are drawn i.i.d. from a common prior distribution $\buyerCDF$. From \cite{myerson1981optimal}, the optimal mechanism maximizes the virtual welfare for regular distributions, and hence 
 \begin{equation}
 \label{eq:myerson-rev}
 \opt{\vbf}=\underset{\mathbf{x}\in[0,1]^{\buyerset}: \sum_{i\in\buyerset}{x_i}\leq m}{\max} \sum_ix_i\virtualV(v_i)~.
 \end{equation}
 In particular, when $m=1$, $\opt{\vbf}=\underset{i\in\buyerset}{\max} \left(\max\{\virtualV(v_i),0\}\right)$. Now, given the optimal expected revenue $\mathbb{E}\left[\opt{\vbf}\right]$ as a benchmark, we are interested in studying the worst-case \emph{revenue approximation ratio} of the best $k$-\dca{} against this benchmark, given by
 \begin{equation}
 \label{eq:comp-ratio}
 \Gamma(k)\triangleq \inf_{\textrm{regular~$\buyerCDF$}}\frac{\underset{\pbf\in\mathbb{R}_{\geq 0}^{\levelNum}:~p_1>p_2>\ldots>p_k}{\max}\mathbb{E}\left[\rev{\vbf}{\pbf}\right]}{\mathbb{E}\left[\opt{\vbf}\right]}.
 \end{equation}

Given the auctioneer's objective, our main goal is to establish a lower bound on $\Gamma(k)$, ideally through a simple and interpretable sequence of bid levels/prices, and study how fast the expected revenue of the best $k$-\dca{} converges to Myerson's optimal revenue.
We start our analysis by considering a seller with a single unit of the item (hence $m=1$) and extend the results to a setting with multiple units in \Cref{sec:multiple}.

\subsection{The Buyers' Bayes-Nash Equilibrium Characterization}
\label{sec:equilibrium}
Considering $k$-\dca{} with i.i.d. buyers as a symmetric game of incomplete information, in the same spirit as the ordinary descending price auction with i.i.d. buyers, we will focus our attention on Bayes-Nash equilibria in which buyers use symmetric strategies; This means a buyer's strategy depends on her valuation, not on her identity. Therefore, the symmetric BNE strategy can be represented by a single bidding function $\eqbid:\mathbb{R}_{\geq 0}\rightarrow\bidset$ that maps each realized value $v$ to one of the possible prices in $\bidset$. Note that our restriction to symmetric BNE is really not a restriction: $k$-\dca{} belongs to the class of symmetric rank-based auctions studied in \cite{chawla2013auctions}. As established in this paper, when buyers' values are i.i.d., any auction in this class has a unique Bayes-Nash equilibrium. Moreover, this equilibrium is symmetric. We next characterize this unique symmetric BNE strategy $\eqbid(v)$. 

Given the decreasing sequence of prices $p_1>\ldots>p_\levelNum$ as possible price levels, a buyer with realized value $v$ chooses an \emph{optimal stopping round} $j$ (if any) along this sequence by bidding the price $p_j$ corresponding to that round, taking as given the strategies of other buyers (at the equilibrium). As a result, in the sequential pricing interpretation of $k$-\dca{}, such a buyer rejects all prices $p_{j'}$ for $j'<j$ and accepts the price $p_j$. If the item is unsold before round $j$, a uniform random buyer among those who have accepted price $p_j$ wins the item.

First, intuitively speaking, higher-valuation buyers are more anxious to purchase than lower-valuation buyers, and hence  buyers with higher valuations accept earlier. As a result, $\eqbid(v)$ should have the form of a monotone increasing step function with discontinuities at certain thresholds $\tau_1\geq\tau_2\geq\ldots\geq\tau_\levelNum>0$, and hence buyers' equilibrium can be represented by this sequence of thresholds. Second, we can actually provide an explicit relationship between the sequence of prices $\mathbf{p}=(p_1, \dots, p_k)$ and the sequence of equilibrium thresholds $\boldsymbol{\tau}=(\tau_1, \dots, \tau_k)$, which gives us our desired characterization of the symmetric BNE.
\begin{proposition}
\label{prop:BNE}
Fix a value distribution $\buyerCDF$. For any given sequence of strictly decreasing prices $\mathbf{p}$, there exists a sequence of thresholds $+\infty\triangleq\tau_0>\tau_1\geq\tau_2\geq\ldots\geq\tau_\levelNum>0$ such that
\begin{enumerate}[label=(\roman*)]
    \item The symmetric Bayes-Nash equilibrium $\eqbid$ for the buyers is to bid $p_j$ (or equivalently stop at the price of round $j$) if their valuation is in $[\tau_j,\tau_{j-1})$ for $j\in[k]$ and bid $0$ (or equivalently never stop at any price) if their valuation is smaller than $\tau_k$.
    \item We have $\tau_k=p_k$ and the thresholds $\tau_1, \dots, \tau_{k-1}$ satisfy 
\begin{align}\label{Eq:Pro:PricesToThresholds-new}
  \frac{1- \left(\frac{\buyerCDF(\tau_j)}{\buyerCDF(\tau_{j-1})} \right)^n}{1- \frac{\buyerCDF(\tau_j)}{\buyerCDF(\tau_{j-1})}} \frac{\tau_{j} - p_j}{n}=\left(\frac{\buyerCDF(\tau_{j})}{\buyerCDF(\tau_{j-1})} \right)^{n-1} \frac{1- \left(\frac{\buyerCDF(\tau_{j+1})}{\buyerCDF(\tau_{j})} \right)^n}{1- \frac{\buyerCDF(\tau_{j+1})}{\buyerCDF(\tau_{j})}} \frac{\tau_{j} - p_{j+1}}{n}, \quad \text{ for } j=1, \dots, k-1.
\end{align}
\end{enumerate}
\end{proposition}
We defer the proof of \Cref{prop:BNE} to \Cref{apx:proof-prop-BNE}. We highlight that \cref{Eq:Pro:PricesToThresholds-new} is the indifference conditions for a buyer with value $\tau_j$: if she purchases at price $p_j$, the left-hand side is her expected utility and if she purchases at price $p_{j+1}$, the right-hand side is her expected utility. Also, if a buyer with value $v$ chooses a stopping round $j$, then we should have $v\geq p_j$ to guarantee her individual rationality. Given our equilibrium characterization above, this property automatically holds as $\tau_j\geq p_j$ for all $j\in[\levelNum]$. The proof is simple and based on backward induction: For $j=k$ this is true as $\tau_k=p_k$. Now assume $\tau_{j+1}\geq p_{j+1}$. Then the right-hand-side of  \cref{Eq:Pro:PricesToThresholds-new} is non-negative as $\tau_j\geq\tau_{j+1}\geq p_{j+1}$. Therefore $\tau_j\geq p_j$, completing the proof of the inductive step. 

Given a value distribution $\buyerCDF$, our characterization defines a mapping $\PriceThresh:\mathbb{R}_{\geq 0}^\levelNum\rightarrow\mathbb{R}_{\geq 0}^\levelNum$ that maps any sequence of strictly decreasing prices $\pbf$ to a sequence of (weakly) decreasing equilibrium thresholds $\boldsymbol\tau$, or equivalently maps a sequence of $k$ distinct prices $p_1>\ldots>p_k$ to a sequence of $k'\leq k$ distinct equilibrium thresholds $\tau_1>\ldots>\tau_{k'}$.\footnote{We highlight that not all price sequences are \emph{minimal}, meaning that their corresponding distinct threshold sequence is of the same length; however, as we see later, our mechanisms always generate minimal prices.}  We next show that $\PriceThresh$ has an inverse, denoted by $\ThreshPrice$, which helps us to translate equilibrium thresholds to prices. We postpone the proof of \Cref{Pro:PriceToThresholds:Surjective-new} to \Cref{apx:proof-of-inverse-map}.

\begin{proposition}\label{Pro:PriceToThresholds:Surjective-new}
For any sequence of $k$ strictly decreasing thresholds $\boldsymbol{\tau}$, there exists a sequence of $k$ strictly decreasing prices $\mathbf{p}=\ThreshPrice(\boldsymbol{\tau})$ such that the corresponding buyers' symmetric Bayes-Nash equilibrium is determined by thresholds $\boldsymbol{\tau}$. 
\end{proposition}

Equipped with Propositions~\ref{prop:BNE} and \ref{Pro:PriceToThresholds:Surjective-new}, the auctioneer can directly work with the sequence of equilibrium thresholds instead of the sequence of prices in the revenue maximization problem. Based on this idea, we next reformulate the auctioneer's problem in terms of an alternative problem which we call \emph{batched prophet inequality}.

\subsection{Reduction to ``Batched Prophet Inequality''}
\label{sec:reduction}
We start this section by defining the following variant of the basic prophet inequality problem \cite{samuel1984comparison}, which is intimately connected to our analysis of the $k$-\dca{}.
\begin{definition}[Batched Prophet Inequality]
\label{def:batched-prophet}
Consider a decision-maker maximizing her expected reward in a sequential game with $k$ rounds. Before the beginning of the game, the decision-maker picks $k$ thresholds $\tau_1>\tau_2>\ldots>\tau_k$. Then $n$ rewards $V_1,\dots, V_n$ are drawn independently from distribution $F$ (known by the decision-maker). The game then starts from round $1$. At each round $i$, if all the rewards are below threshold $\tau_i$, the game advances to next round $i+1$. Otherwise, the decision-maker wins one of the rewards that passes threshold $\tau_i$ uniformly at random and the game ends. If no reward passes any of thresholds until the end of round $k$, the game ends with decision-maker winning zero reward.
\end{definition}

If the decision-maker knows the reward realizations, the expected optimal reward is: $$\maxV\triangleq\mathbb{E}\left[\max\left\{\max_{i \in [n]} V_i,0\right\}\right]~.$$
Note that $V_i$'s are allowed to be negative, but the decision-maker has always the option of rejecting any negative rewards by choosing $\tau_k\geq 0$. We are interested in designing thresholds $\boldsymbol{\tau}$ to maximize the  ratio of the expected reward of the decision-maker, denoted by $\algT(\tau_1,\ldots,\tau_k)$, to the offline benchmark $\maxV$. This ratio is known as the \emph{competitive ratio}:
\begin{align}
\frac{\algT(\tau_1,\ldots,\tau_k)}{\maxV}.
\end{align}

To make a connection between the problem of maximizing the competitive ratio in the batched prophet inequality setting and the revenue approximation ratio of $k$-\dca{} versus the Myerson's optimal mechanism (defined in \cref{eq:comp-ratio}), we rely on two key observations:
\begin{itemize}
    \vspace{1mm}
    \item We evaluate the expected revenue of our $k$-\dca{} at its symmetric BNE. As a result, we can rely on Myerson's payment/revenue equivalence lemma~\citep{myerson1981optimal} to simplify our analysis: for regular distributions, the expected revenue of $k$-\dca{} is equal to the expected virtual value of the winner. 
    \vspace{1mm}
    \item Suppose the auctioneer selects a sequence of equilibrium thresholds $\hat{\tau}_1>\ldots>\hat{\tau}_k$ (as in \Cref{prop:BNE}), which can always be induced by a sequence of prices $p_1>\ldots>p_k$, where $\pbf=\ThreshPrice(\boldsymbol{\hat{\tau}})$ (as in \Cref{Pro:PriceToThresholds:Surjective-new}). Then we can \emph{simulate} the winner-selection process of $k$-\dca{} by finding the first round $j$ in which one of the buyers $v_1,\ldots,v_n$ passes threshold $\hat{\tau}_j$, and picking one such buyer uniformly at random. 
\end{itemize}
Now consider an instance of the batched prophet inequality where $V_i=\virtualV(v_i)$ --- hence, $F$ is the distribution of the random variable $\virtualV(v)$ for $v\sim\buyerCDF$, or equivalently $F(x)=G\left(\virtualV^{-1}(x)\right)$. Suppose the decision-maker selects thresholds $\tau_1>\ldots>\tau_k>0$. Also, consider running $k$-\dca{} with equilibrium thresholds $\hat{\tau}_1>\ldots>\hat{\tau_k}$ against buyers with i.i.d. values $\vbf$ drawn from $\buyerCDF$. Then the expected reward of the decision-maker is equal to the expected virtual value of the winner of the $k$-\dca{}, if any, (and hence equal to its expected revenue), if and only if $\tau_i=\virtualV(\hat{\tau}_i)$ for all $i\in[k]$. This claim simply holds as  $\virtualV$ is increasing due to regularity. Note that $\virtualV(\hat{\tau}_k)=\tau_k>0$, and hence no buyer with a negative virtual value can ever be a winner in the resulting $k$-\dca{}. Now, the offline benchmark of the batched prophet inequality is the same as  the expected revenue of Myerson's optimal auction, i.e.,
$$
\maxV=\mathbb{E}\left[\max\left\{\max_{i \in [n]} V_i,0\right\}\right]=\mathbb{E}\left[\max\left\{\max_{i \in [n]} \virtualV(v_i),0\right\}\right]=\mathbb{E}\left[\opt{\vbf}\right]~.
$$
The final step is translating thresholds $\hat{\tau}_i=\virtualV^{-1}(\tau_i)$ to prices. This can be done using \cref{Pro:PriceToThresholds:Surjective-new}, which results in $\pbf=\ThreshPrice\left([\hat{\tau}_i]_{i\in\buyerset}\right)$. Putting all pieces together, the revenue approximation of $k$-\dca{} with prices $\pbf$ is equal to the competitive ratio of thresholds $\boldsymbol{\tau}$ picked by the decision-maker in the above batched prophet inequality instance. 

We end this section by showing how the problem of finding optimal sequence of thresholds in batched prophet inequality can be reformulated as a simple dynamic programming.

\vspace{3mm}
\noindent\textbf{{{Dynamic Programming:}}}~~~  Initially the decision-maker only knows the prior distribution of the rewards and nothing more about their realizations. Yet, the decision-maker knows when the game reaches round $j$, her information about the reward distribution is going to change. In particular, if round $j$ has arrived and no reward is collected, then the decision-maker should know all the past rewards are smaller than $\tau_{j-1}$, and her posterior belief about the rewards would change to the conditional CDF $\frac{F(v)}{F(\tau_{j-1})}$ over the support $[0, \tau_{j-1})$. Therefore, the state of the system at any round is the remaining number of rounds and the current upper bound on the distribution of rewards (i.e., the lowest threshold so far). We denote by
\begin{align*}
\Psi(t, \theta)    
\end{align*}
the optimal expected reward if $t$ rounds are remaining and all the rewards are known to be smaller than $\theta$. The following is the Bellman update equation for computing the optimal expected reward starting from state $(t,\theta)$:
\begin{align}\label{eq:Bellman}
    \Psi(t, \theta) = \max_{ \theta' \in [0, \theta]} \Bigg\{& \left( \frac{F(\theta')}{F(\theta)} \right)^n \Psi(t-1, \theta') 
    + \left(1- \left( \frac{F(\theta')}{F(\theta)} \right)^n \right) \mathbb{E}\left[ V \mid  V\in[\theta',\theta) \right] \Bigg\},
\end{align}
where $V\sim F$. We note once a threshold $\theta'$ is picked at state $(t,\theta)$, if the set of rewards $S\subseteq[n]$ passing $\theta'$ is  non-empty,  then the conditional expected collected reward is equal to:
$$
\mathbb{E}\left[\frac{\sum_{i\in S}{v_i}}{\lvert S\rvert}~\big{|}~j\in S: v_j\in[\theta',\theta), j\notin S: v_j<\theta)\right]=\frac{\sum_{i\in S}\mathbb{E}\left[v_i \mid v_i\in[\theta',\theta)\right]}{\lvert S \rvert }=\mathbb{E}\left[ V \mid  V\in[\theta',\theta)\right]~.
$$
which is used in the term corresponding to instantaneous reward in \cref{eq:Bellman}. The recursion in \cref{eq:Bellman} highlights the tradeoff that the decision-maker is facing: by increasing $\theta'$, the probability of collecting instantaneous reward (i.e., $1- \left( \frac{F(\theta')}{F(\theta)} \right)^n $) decreases while the instantaneous expected reward (i.e., $\mathbb{E}\left[ V \mid  V\in[\theta',\theta) \right]$), the probability of future rewards (i.e., $\left( \frac{F(\theta')}{F(\theta)} \right)^n$) and the expected future reward (i.e., $\Psi(t-1, \theta')$) increases. 

Given the above formulation, the optimal expected reward obtained by $k$ thresholds (and the thresholds themselves) can be evaluated by computing $\Psi(k, \infty)$ recursively. However, it is not clear how well these thresholds can approximate the optimum offline reward  $\maxV$
as a benchmark. We address this question in the next section.

\section{Batched Prophet Inequality for a Single Item}
\label{sec:single-unit}

In this section, we focus on the single item batched prophet inequality problem (\Cref{def:batched-prophet}). We first introduce a simple sequence of thresholds that geometrically span the quantile space (\Cref{sec:main-single-item}), and show they provide a competitive ratio against $\maxV$ as a function of $k$ that converges to $1$ exponentially fast (\Cref{sec:thm-sketch}). We further show how to use this result to design price sequences/bid levels in a $k$-\dca ~to obtain approximations to revenue and welfare (\Cref{sec:final-price}). As neither the decision-maker nor the offline benchmark ever accept a negative reward, we assume W.L.O.G that all values in support of $F$ are non-negative in \Cref{sec:main-single-item} and \Cref{sec:final-price}. We revisit this subtle point in \Cref{sec:final-price} when we design our final prices (as virtual values can be negative). 
\subsection{Approximations Using Balanced Thresholds}
\label{sec:main-single-item}

As the main result of this section, we establish that there exists a sequence of $k$ thresholds for the decision-maker that achieves $1-1/e^k$ of the optimum offline reward $\maxV$ in the batched prophet inequality problem. Here, we assume rewards are non-negative.

\begin{theorem}\label{Thm:Comp:ratio}
The following sequence of thresholds achieves $1-1/e^k$ of the optimum offline reward $\maxV$ as the expected reward of the decision-maker in the batched prophet inequality problem:
\begin{align*}
    \tau_1= F^{-1} \left( \left(\frac{1}{e}\right)^{\frac{1}{n}}\right), \tau_2= F^{-1} \left( \left(\frac{1}{e^2}\right)^{\frac{1}{n}}\right), \dots, \tau_k= F^{-1} \left( \left(\frac{1}{e^k}\right)^{\frac{1}{n}}\right).
\end{align*}
\end{theorem}
\begin{remark}
For the special case of $k=1$, our problem boils down to designing a static threshold for the i.i.d. prophet inequality problem~\citep{correa2017posted} or prophet secretary problem (with homogeneous buyers)~\citep{esfandiari2017prophet}. As it has been established in the prior work, no static policy can obtain a competitive ratio better than $1-1/e$, and hence our result is tight for $k=1$. As we show next, our analysis is also tight for general $k$. 
\end{remark}

\vspace{3mm}
\noindent\textbf{{Warm Up (Uniform Distribution):}}~~~ Before providing the proof sketch, let us show how the bound $1- 1/e^k$ appears in an example with uniform distribution over $[0,1]$. Letting 
\begin{align*}
    \alpha=\left(\frac{1}{e}\right)^{\frac{1}{n}},
\end{align*}
the thresholds prescribed in Theorem \ref{Thm:Comp:ratio} become 
\begin{align*}
    \tau_j= \alpha^j \text{ for } j=1, \dots, k.
\end{align*}
The expected reward of the decision-maker by using these thresholds becomes
\begin{align}\label{eq:simple:proof:uniform}
    \algT\left(\tau_1, \dots, \tau_k\right)=\sum_{j=1}^k \left( \alpha^{(j-1)n}- \alpha^{jn} \right) \frac{1}{2}\left(\alpha^{j-1}+ \alpha^{j} \right).
\end{align}
This is because with probability $\alpha^{(j-1)n}- \alpha^{jn}$ the maximum of rewards $\{V_i\}_{i\in[n]}$ falls into interval $[\alpha^{j}, \alpha^{j-1})$. In this case, the game ends by the end of round $j$, while the expected collected reward conditioned on reaching to round $j$ is $\frac{1}{2}\left(\alpha^{j-1}+ \alpha^{j} \right)$. We can rewrite \eqref{eq:simple:proof:uniform} as
\begin{align}\label{eq:simple:proof:uniform:algorithm}
    \frac{1}{2} \left(1- \alpha^n \right) \left(1+ \alpha \right) \sum_{j=0}^{k-1} \alpha^{j (n+1)} =& \frac{1}{2} \left(1- \alpha^n \right) \left(1+ \alpha \right) \left( \frac{1-\alpha^{(n+1)k}}{1- \alpha^{n+1}} \right) \nonumber \\
    & = \frac{1}{2} \left(1- \frac{1}{e} \right) \left( 1+ \frac{1}{e^{1/n}} \right) \frac{1- \frac{1}{e^{k+ k/n}}}{1- \frac{1}{e^{1+ 1/n}}},
\end{align}
where the last equality follows by plugging in $\alpha=\left(\frac{1}{e}\right)^{\frac{1}{n}}$. The optimum offline reward $\maxV$, on the other hand, is equal to
\begin{align}\label{eq:simple:proof:uniform:optimal}
    \maxV=\mathbb{E} \left[ \max_{i \in \mathcal{N}} V_i \right] = \int_{0}^1 \left(1- z^n \right) dz = 1- \frac{1}{n+1}.
\end{align}
We next compare the performance of our thresholds given in \eqref{eq:simple:proof:uniform:algorithm} to the optimal offline reward given \eqref{eq:simple:proof:uniform:optimal}. We have
\begin{align*}
  \algT(\tau_1,\ldots,\tau_k)= \frac{1}{2} \left(1- \frac{1}{e} \right) \left( 1+ \frac{1}{e^{1/n}} \right) \frac{1- \frac{1}{e^{k+ k/n}}}{1- \frac{1}{e^{1+ 1/n}}} \ge & \lim_{n \to \infty} \frac{1}{2} \left(1- \frac{1}{e} \right) \left( 1+ \frac{1}{e^{1/n}} \right) \frac{1- \frac{1}{e^{k+ k/n}}}{1- \frac{1}{e^{1+ 1/n}}} \\
   = & 1- \frac{1}{e^k} \ge \left(1- \frac{1}{e^k} \right)\maxV,
\end{align*}
establishing the $1- \frac{1}{e^k}$ competitive ratio against the optimum offline reward as a benchmark.

\subsection{Proof Sketch of Theorem \ref{Thm:Comp:ratio} for General Distributions}
\label{sec:thm-sketch}
Here, we provide the proof sketch of \Cref{Thm:Comp:ratio:revenue} and relegate the details to \Cref{apx:proof-thm-comp-ratio}. By choosing the first threshold $\tau_1$, we obtain at least a reward of $\tau_1$ if at least one of the random variables is above this threshold. In addition to $\tau_1$, we obtain the difference between the random variable $V_i$ that is selected (if any) and the threshold $\tau_1$, i.e., $\mathbb{E}[(V_i-\tau_1)^+]$. There is a chance that this selected random variable is the highest random variable and by bounding this probability we establish that the expected instantaneous reward,  i.e., expected reward obtained by selecting the first threshold, is at least
\begin{align}\label{eq:firstround}
    \left(1- F(\tau_1)^n \right) \tau_1 + P_n(F(\tau_1)) \sum_{i=1}^n \mathbb{E}[(V_i-\tau_1)^+],
\end{align}
where for any $x \in [0,1]$ we define the polynomial $P_n(x)$ as 
\begin{align}\label{eq:Polynomial}
    P_n(x)\triangleq \sum_{i=0}^{n-1} \frac{1}{i+1} \binom{n-1}{i} x^{n-1-i} (1-x)^{i}.
\end{align}
\Cref{eq:firstround} manifests the first tradeoff that the decision-maker is facing: by selecting a threshold $\tau_1$, she balances the terms $\tau_1$ and $\mathbb{E}[(V_{\mathrm{max}}-\tau_1)^+]$, which are increasing and decreasing in $\tau_1$, respectively. 

To gain some intuitions, let us first consider the simpler case of $k=1$. If we only had one round, the decision-maker could safely aim to only maximize the instantaneous reward (as there are no future rounds). To this end, the optimal threshold should make both coefficients $(1- F(\tau_1)^n)$ and $P_n(\tau_1)$ large. We show that it is possible to have 
\begin{align}\label{eq:bound:min}
    \min\{(1-F(\tau_1)^n), P_n(F(\tau_1))\} \ge 1- \frac{1}{e}.
\end{align}
In particular, for $F(\tau_1)=1-\frac{1}{n}$ the above inequality holds. We can then use the fact that  $$\tau_1+\sum_{i=1}^{n}\mathbb{E}[(V_i-\tau_1)^+]\geq \tau_1+\mathbb{E}[(V_{\mathrm{max}}-\tau_1)^+]\geq \maxV~,$$
 establishing the $1- \frac{1}{e}$ competitive ratio for $k=1$. In fact, this result is closely related to the well-known Bernoulli selection lemma (see \cite{correa2017posted,esfandiari2017prophet}) and online contention resolution schemes for i.i.d. variables under random order (see \cite{yan2011mechanism,lee2018optimal}). These techniques also lead to the well-known result that there exits a static threshold for the i.i.d. prophet inequality problem that obtains the approximation ratio  $1-1/e$, and our analysis for this special cases provides an alternative proof for it.



However, we are interested to approximate the expected reward for any $k \ge 1$ number of rounds, where the decision-maker is allowed to use $k$ different thresholds. Here, the decision-maker not only should try to keep the instantaneous reward high, she should also hedge against the future and have an eye on the expected reward of future rounds. 

To guide the analysis, let us consider the expected reward of the second round. With probability $F(\tau_1)^n$, all random variables are below $\tau_1$ and we get to the second round. The stage-reward of the second round is the same as the first round by replacing the distribution $F(x)$ with $\frac{F(x)}{F(\tau_1)}$, which is basically the distribution of $V_i$ conditioned on $V_i<\tau_1$. We can write  
\begin{align}\label{eq:pf:bound:explain}
   \mathbb{E}[\algT(\tau_1, \dots, \tau_k)]
    \ge & \left(1- F(\tau_1)^n \right) \tau_1 + P_n(F(\tau_1)) \sum_{i=1}^n \mathbb{E}[(V_i-\tau_1)^+] \nonumber \\
    & + F(\tau_1)^n  \mathbb{E}\left[ \algT(\tau_1, \dots, \tau_k) \mid V_j \le \tau_1, j \in [n] \right].
\end{align}
\Cref{eq:pf:bound:explain} manifests the second tradeoff that the decision-maker is facing: by choosing $\tau_1$ she needs to make $\min\{(1-F(\tau_1)^n), P_n(F(\tau_1))\}$ large, but crucially she needs to make the term $F(\tau_1)^n$ large enough at the same time. We next show in the following lemma that it is possible to satisfy \eqref{eq:bound:min} while having $F(\tau_1)^n \ge \frac{1}{e}$. We postpone its proof to \Cref{apx:lemma:balance}.
\begin{lemma}\label{Lem:inequality}
For $x=\left(\frac{1}{e}\right)^{1/n}$, we have 
\begin{align*}
    \min\left\{1- x^n, P_n(x)\right\} \ge  1-\frac{1}{e} \quad \text{ for all } n.
\end{align*}
\end{lemma}
Using Lemma \ref{Lem:inequality}, we can further bound \eqref{eq:pf:bound:explain} as 
\begin{align*}
    & \left(1- \frac{1}{e} \right) \left(\tau_1 + \sum_{i=1}^n \mathbb{E}[(V_i-\tau_1)^+] \right) + \frac{1}{e} \left( \left(1- \frac{1}{e} \right) \left(\tau_2 + \sum_{i=1}^n \mathbb{E}[(V_i-\tau_2)^+] \right)+ \dots \right)  \\
    & \ge \left(1- \frac{1}{e} \right) \mathrm{OPT} + \frac{1}{e} \left( \left(1- \frac{1}{e} \right) \mathrm{OPT} + \frac{1}{e} \cdots   \right) = \mathrm{OPT} \left(1- \frac{1}{e^k} \right).
\end{align*}
which again uses the fact that $\tau + \sum_{i=1}^n \mathbb{E}[(V_i-\tau)^+]\geq \maxV$ for any $\tau$, completing the proof. 

\subsection{The Price Trajectories for Revenue and Welfare Approximations}
\label{sec:final-price}
\Cref{Thm:Comp:ratio} finds an approximately optimal sequence of $k$ distinct thresholds in the batched prophet inequality setting. Using the reduction of \Cref{sec:reduction}, we can determine the equilibrium thresholds of our final $k$-level descending price auction, as well as its set of $k$ distinct prices/bid levels supporting this equilibrium. The only subtle difference is that $V_i=\virtualV(v_i)$ can take negative values when $v_i<\rho\triangleq \virtualV^{-1}(0)$. However, neither the decision-maker nor the optimal offline benchmark should accept any negative $V_i$ (which is equivalent to allocating to a buyer with negative virtual value). We handle this subtlety by basically focusing on the subset of buyers for which $v_i\geq \rho$, and constructing our thresholds by invoking \Cref{Thm:Comp:ratio} for the conditional distribution $$\bar{F}(x)=\mathbb{P}\left[V_i\leq x | V_i\geq 0\right]=\frac{F(x)-F(0)}{1-F(0)}=\frac{G\left(\virtualV^{-1}(x))\right)-G(\rho)}{1-G(\rho)}.$$

We summarize our procedure for constructing these prices in Algorithm~\ref{alg:prices-revenue}.
\begin{algorithm}[htb]
 	\caption{Price construction of $k$-\dca ~for approximating optimal revenue}
 	\label{alg:prices-revenue}
 	\KwIn{number of distinct prices $k\in\mathbb{N}$, buyers' (regular) value distribution $\buyerCDF$}
 	\KwOut{sequence of prices $p_1>p_2>\ldots>p_k$}
 	\vspace{1mm}
 	Define $\rho\triangleq \virtualV^{-1}(0)$ and $\alpha\triangleq\frac{1}{e^{1/n}}$\\
 	
 	\For{$j=1,\ldots,k$}{
 	Let $\hat\tau_j=G^{-1}\left(\left(1-G(\rho)\right)\alpha^j+G(\rho)\right)$\\
 	}
 	Define $\beta_j\triangleq G(\hat\tau_j) $ for $j=1,\ldots,k$ and $\beta_0\triangleq 1$\\
 	Let $p_k=\hat\tau_k$\\
 	\For{$j=k-1,\ldots,1$}{
 	 $p_j=\hat\tau_j- \left(\hat\tau_j-p_{j+1}\right)\left(\frac{{\beta_j}^n-{\beta_{j+1}}^n}{\beta_j-\beta_{j+1}}\right)\left(\frac{\beta_{j-1}-\beta_{j}}{{\beta_{j-1}}^n-{\beta_{j}}^n}\right)$
 	}
\end{algorithm}

\begin{theorem}\label{Thm:Comp:ratio:revenue}
The $k$-level descending price auction for selling a single item to i.i.d. buyers, with prices constructed by Algorithm~\ref{alg:prices-revenue}, achieves an expected revenue no less than $1-\frac{1}{e^k}$ of the optimal revenue.
\end{theorem}

We defer the proof of \Cref{Thm:Comp:ratio:revenue} to \Cref{apx:proof-thm-revenue}.

\begin{remark}
\label{rem:price-welfare}
So far the focus of our paper has been on maximizing expected virtual welfare, which is equivalent to maximizing expected revenue. However, the same approach can help us to find a sequence of prices, so that the expected value of the winner of the $k$-\dca~ approximates the expected maximum social welfare, i.e., $\mathbb{E}\left[\max_{i\in\buyerset}v_i\right]$. To this end, we only need to define an instance of the batched prophet inequality problem where $V_i=v_i$, and then use the thresholds $\boldsymbol{\tau}$ in \Cref{Thm:Comp:ratio} as equilibrium thresholds (assuming $V_i\sim G$). Combining \Cref{Pro:PriceToThresholds:Surjective-new} and \Cref{Eq:Pro:PricesToThresholds-new} with the fact that
\begin{align*}
    G(\tau_j)=\frac{1}{e^{j/n}} \text{ for }  j=1,\ldots,k
\end{align*}
results in a sequence of prices $\pbf$ satisfying:
\begin{align}\label{eq:Pro:thresholds:unifomr}
   p_j=\frac{1}{e^{j/n}} \left( 1- \frac{1}{e^{(n-1)/n}} \right) + \frac{1}{e^{(n-1)/n}} p_{j+1}~~\text{for}~~j=1, \dots, k-1,
\end{align}
with the initialization $p_k=G^{-1}\left(\frac{1}{e^{k/n}}\right)$. The proof is similar to (and somewhat simpler than) that of \Cref{Thm:Comp:ratio:revenue} and omitted for brevity.
\end{remark}

\begin{figure}[t]
     \centering
     \begin{subfigure}[b]{0.45 \textwidth}
         \centering
         \includegraphics[width= \textwidth]{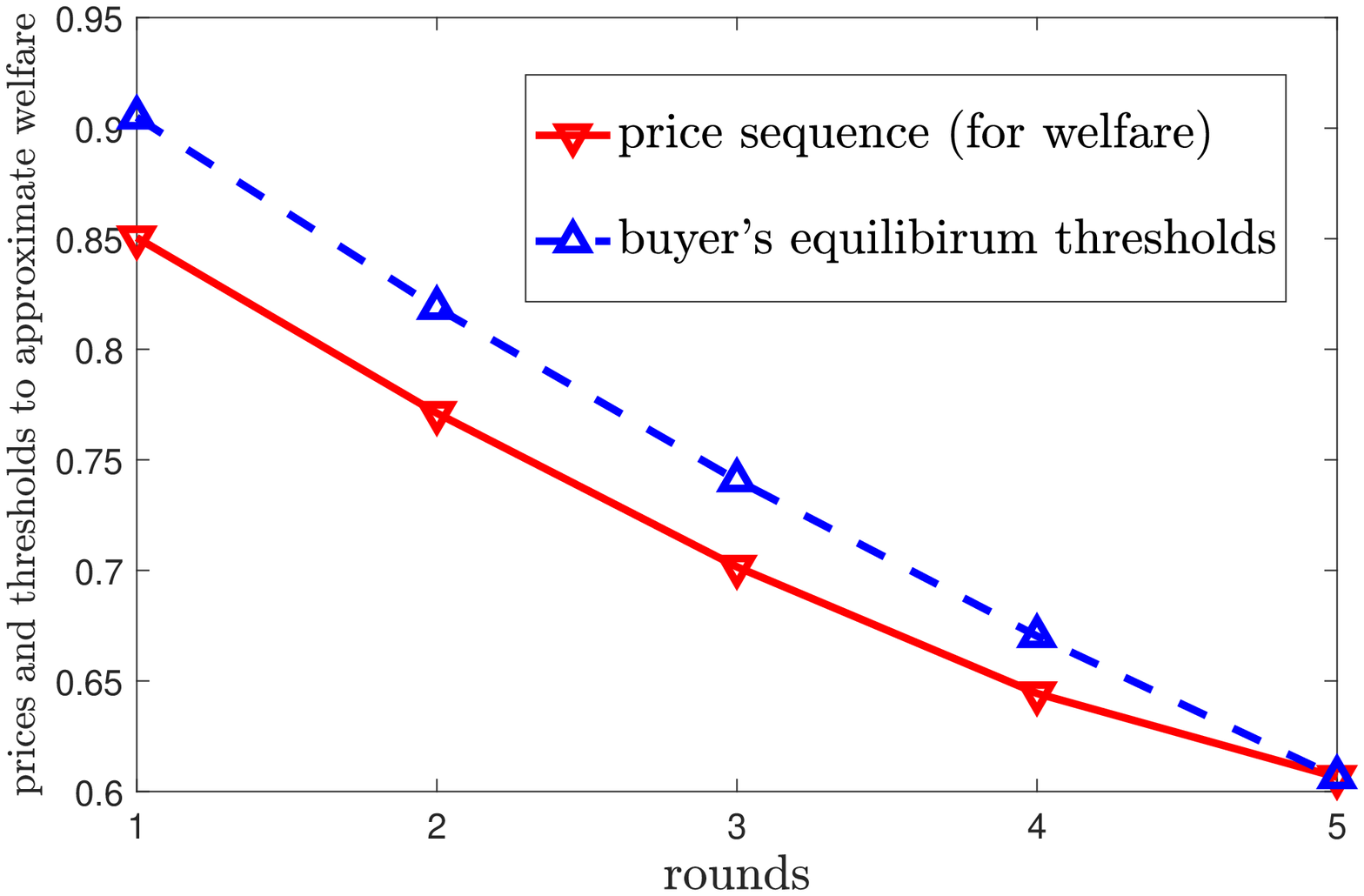}
         \caption{}
         \label{fig:central:variance}
     \end{subfigure}
     \hfill
     \begin{subfigure}[b]{0.45 \textwidth}
         \centering
         \includegraphics[width=\textwidth]{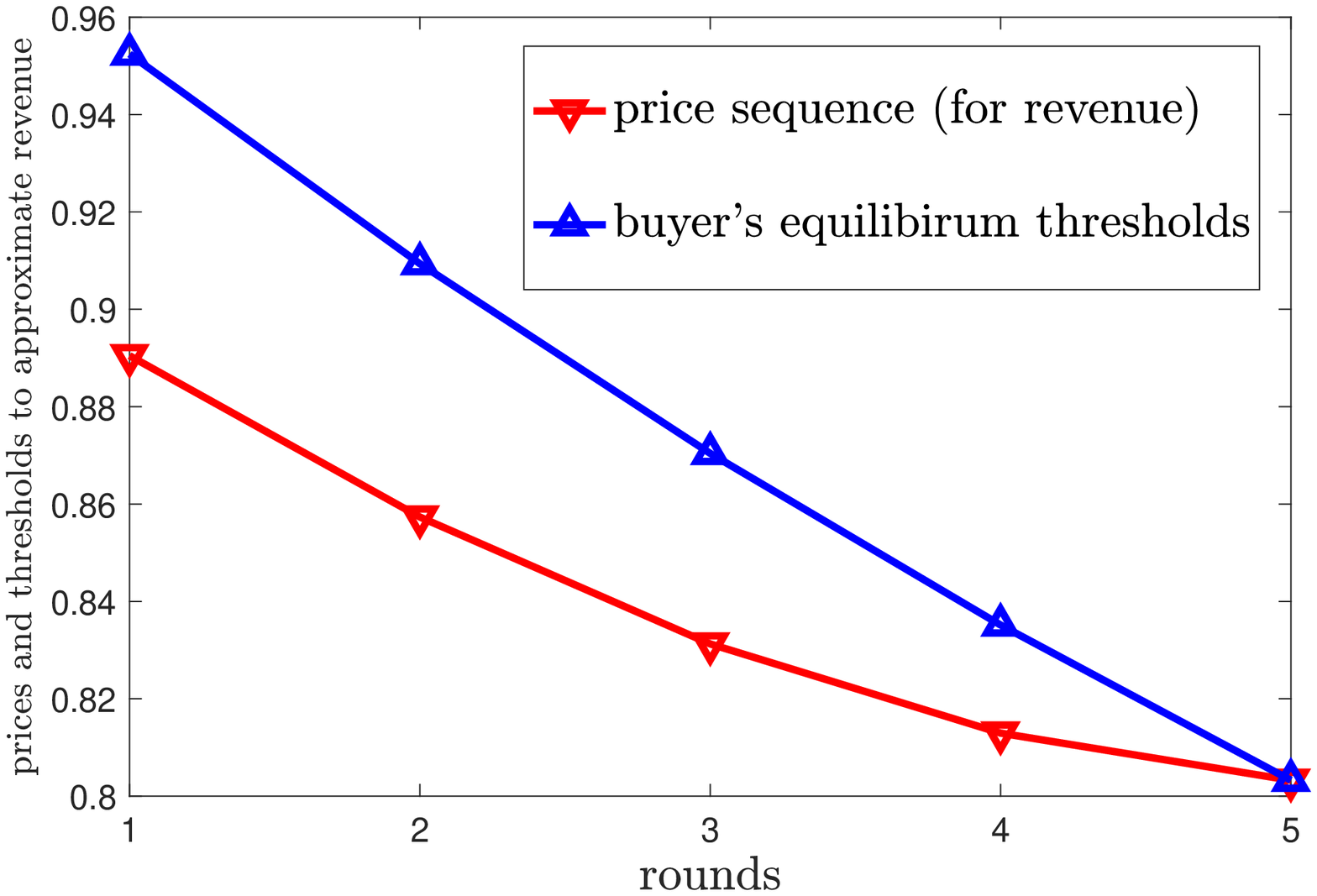}
         \caption{}
         \label{fig:local:variance}
     \end{subfigure}
    \caption{The sequence of prices and the corresponding thresholds that determine the equilibrium for uniform distribution over $[0,1]$, $n=10$ buyers, and $k=5$ rounds: (a) the prices and the corresponding thresholds that determine the buyer's equilibrium for approximating welfare and (b) the prices and the corresponding thresholds that determine the buyer's equilibrium for approximating revenue }\label{Fig:example:uniform}
\end{figure}

Figure \ref{Fig:example:uniform} illustrates the price and the equilibrium trajectories described in \Cref{Thm:Comp:ratio:revenue} and \Cref{rem:price-welfare} for an example with $n=10$ buyers whose values are drawn from uniform distribution over $[0,1]$, and with $k=5$ price levels for approximating the optimal welfare and the optimal revenue. We observe that in the last round the prices and the equilibrium threshold coincides, but for any of the early rounds the buyers purchase only when their value is larger than a threshold which is strictly larger than the price. This is because of the competition among buyers: by waiting, or equivalently settling for a lower price in the auction,  even though buyers face a lower price, the competition among buyers increases and the chances of acquiring the item decreases.

\section{Extension to Multiple Items}
\label{sec:multiple}
So far we considered a setting in which the seller has a single item. In this section, we extend our results to a setting with $m \ge 2$ identical items and $n$ unit-demand buyers.
\subsection{The environment}
 As in our base model, the seller announces a sequence of $k$ prices and the buyers simultaneously decide the price at which they accept to purchase the item. Similar to this baseline model, for any descending sequence of prices $\mathbf{p}$, there exists a sequence of thresholds $\tau_1\geq\tau_2\geq\ldots\geq\tau_\levelNum>0$ that characterize the buyer's equilibirum:

\begin{proposition}
\label{Pro:PricesToThresholds:multipleitems}
Fix a value distribution $\buyerCDF$. For any given sequence of strictly decreasing prices $\mathbf{p}$, there exists a sequence of thresholds $+\infty\triangleq\tau_0>\tau_1\geq\tau_2\geq\ldots\geq\tau_\levelNum>0$ such that the symmetric Bayes-Nash equilibrium $\eqbid$ for the buyers is to bid $p_j$ (or equivalently stop at the price of round $j$) if their valuation is in $[\tau_j,\tau_{j-1})$ for $j\in[k]$ and bid $0$ (or equivalently never stop at any price) if their valuation is smaller than $\tau_k$.
\end{proposition}
This proposition is the analogue of Proposition \ref{prop:BNE} for multiple items. In the proof of this proposition, given in the appendix, we also provide an explicit characterization of the equilibrium thresholds in terms of the sequence of prices. 

\subsection{Batched Prophet Inequality for Multiple Items}
Similar to our baseline analysis for a single item, if the decision-maker knows the reward realizations, the expected optimal reward becomes  $$\maxV_{m}\triangleq\mathbb{E}\left[\max_{0\le \ell \le m}\sum_{i=1}^\ell V_{(i)}\right],$$
where $V_{(i)}$ is the $i$-th top value and by convention for $\ell=0$ the reward is zero. This expression is the expectation of the total reward when the decision-maker can take up to $m$ rewards. We are interested in designing thresholds $\boldsymbol{\tau}$ to maximize the  ratio of the expected reward of the decision-maker, denoted by $\algT_m(\tau_1,\ldots,\tau_k)$, to the offline benchmark $\maxV$. That is the competitive ratio, given by 
\begin{align}
\frac{\algT_m(\tau_1,\ldots,\tau_k)}{\maxV_m}.
\end{align}



\subsection{Approximations Using Balanced Thresholds for Multiple Items}
In this section, we devise a sequence of $k$ thresholds
for the decision-maker and establish its approximation of the optimum offline reward $\maxV_m$ in the
batched prophet inequality problem with $m>1$ items.
\begin{theorem}\label{Thm:Comp:ratio:welfare:multiple}
For any $m \ge 2$ items and $k \ge 1$ rounds, there exists $N(\epsilon)$ such that the sequence of thresholds  
\begin{align*}
    \tau_r= F^{-1}\left(\left( 1-\frac{m}{n}\right)^r \right) \quad \text{ for } r=1, \dots, k
\end{align*}
for $n \ge N(\epsilon)$ achieves
\begin{align*}
     \left(1- e^{-m} \frac{m^m}{m!} \right) + \sum_{r=1}^{k-1} \sum_{i=0}^{m-1} \frac{m^{i} r^{i}}{i!} \frac{e^{-mr}}{1+\epsilon} \left(1- e^{-(m-i)} \frac{(m-i)^{m-i}}{(m-i)!} \right)
\end{align*}
of the optimum offline reward $\maxV_m$ as the expected reward of a decision-maker that can acquire up to $m$ rewards in the batched prophet inequality.
\end{theorem}
In the rest of this subsection, we provide the proof of this theorem.

For any $\tau$, we let $$
    S^+(\tau)=\sum_{i=1}^m \mathbb{E}[(V_{(i)}-\tau)^+]$$ where $V_{(i)}$ is the top $i$-th random variable. We also define the following polynomials:
\begin{align*}
    A(n, m, x) = &\frac{1}{m} \sum_{i=1}^n x^{n-i} (1-x)^i \binom{n}{i} \min\{i, m\} \\
    B(n, m, x) = & \sum_{i=0}^{n-1} x^{n-1-i} (1-x)^i \binom{n-1}{i} \min\left\{1, \frac{m}{i+1}\right\}.
\end{align*}

Using these notations, we can write the expected welfare as
\begin{align}\label{app:eq:pf:multipleitems:1}
    &m \tau_1 A(n, m, F(\tau_1)) + S^+(\tau_1) B(n, m, F(\tau_1)) \nonumber \\
    &+ \sum_{j_1=0}^{m-1} \left( 1- F(\tau_1)\right)^{j_1} F(\tau_1)^{n-j_1}  \Biggl( m \tau_2 A\left(n-j_1, m-j_1, \frac{F(\tau_2)}{F(\tau_1)}\right) + S^+(\tau_2) B\left(n-j_1, m-j_1, \frac{F(\tau_2)}{F(\tau_1)}\right) \nonumber \\
    &+ \sum_{j_2=0}^{m-j_1-1} \binom{n-j_1}{j_2} \left( 1- \frac{F(\tau_2)}{F(\tau_1)}\right)^{j_2} \left( \frac{F(\tau_2)}{F(\tau_1)}\right)^{n-j_1-j_2} \nonumber \\
    & \Biggl(m \tau_3 A\left(n-j_1-j_2, m-j_1-j_2, \frac{F(\tau_3)}{F(\tau_2)}\right) + S^+(\tau_3) B\left(n-j_1-j_2, m-j_1-j_2, \frac{F(\tau_3)}{F(\tau_2)}\right) \nonumber \\
    &  \vdots \nonumber \\
    & + m \tau_k A\left(n-\sum_{\ell=1}^{k-1}j_{\ell}, m-\sum_{\ell=1}^{k-1}j_{\ell}, \frac{F(\tau_{k})}{F(\tau_{k-1})}\right) + S^+(\tau_{k}) B\left(n-\sum_{\ell=1}^{k-1}j_{\ell}, m-\sum_{\ell=1}^{k-1}j_{\ell}, \frac{F(\tau_{k})}{F(\tau_{k-1})}\right) \nonumber\\
    & \overset{(a)}{=} m \tau_1 A(n, m, \tau_1) + S^+(\tau_1) B(n, m, F(\tau_1)) \nonumber \\
    & +\sum_{i_1=0}^{m-1} \binom{n}{i_1} \left( 1- F(\tau_1)\right)^{i_1} F(\tau_1)^{n-i_1}  \left( m \tau_2 A\left(n-i_1, m-i_1, \frac{F(\tau_2)}{F(\tau_1)}\right) + S^+(\tau_2) B\left(n-i_1, m-i_1, \frac{F(\tau_2)}{F(\tau_1)}\right) \right) \nonumber\\
    & +\sum_{i_2=0}^{m-1} \binom{n}{i_2} \left( 1- F(\tau_2)\right)^{i_2} F(\tau_2)^{n-i_2}  \left( m \tau_2 A\left(n-i_2, m-i_2, \frac{F(\tau_3)}{F(\tau_2)}\right) + S^+(\tau_3) B\left(n-i_1, m-i_1, \frac{F(\tau_3)}{F(\tau_2)}\right) \right) \nonumber \\
    & \vdots \nonumber \\
    & +\sum_{i_{k-1}=0}^{m-1} \binom{n}{i_{k-1}} \left( 1- F(\tau_k)\right)^{i_{k-1}} F(\tau_k)^{n-i_{k-1}} \nonumber \\
    & \left( m \tau_k A\left(n-i_{k-1}, m-i_{k-1}, \frac{F(\tau_k)}{F(\tau_{k-1})}\right) + S^+(\tau_3) B\left(n-i_{k-1}, m-i_{k-1}, \frac{F(\tau_k)}{F(\tau_{k-1})}\right) \right),
\end{align}
where (a) follows from the following argument. For any $r \in \{2, \dots, k\}$, the coefficient of $$\left( m \tau_{r} A\left(n-\sum_{\ell=1}^{r-1}j_{\ell}, m-\sum_{\ell=1}^{r-1}j_{\ell}, \frac{F(\tau_{r})}{F(\tau_{r-1})}\right) + S^+(\tau_{r}) B\left(n-\sum_{\ell=1}^{r-1}j_{\ell}, m-\sum_{\ell=1}^{r-1}j_{\ell}, \frac{F(\tau_{r})}{F(\tau_{r-1})}\right) \right)$$ where $\sum_{\ell=1}^{r-1}j_{\ell}=i_r$ is 
\begin{align*}
    \sum_{j_1, \dots, j_{r-1} \in \{1, \dots, m-1\}, \sum_{\ell=1}^{r-1} j_{\ell}=i_r} & \binom{n}{j_1} \left( 1- F(\tau_1)\right)^{j_1} F(\tau_1)^{n-j_1} \binom{n-j_1}{j_2} \left( 1- \frac{F(\tau_2)}{F(\tau_1)}\right)^{j_2} \left(\frac{F(\tau_2)}{F(\tau_1)}\right)^{n-j_2} \\
    &\cdots  \binom{n-j_1 - \dots - j_{k-1}}{j_k} \left( 1- \frac{F(\tau_{k})}{F(\tau_{k-1})}\right)^{j_k} \left( \frac{F(\tau_{k})}{F(\tau_{k-1})} \right)^{n-j_1- \cdots - j_k} \\
    & = \binom{n}{i_r} \left( 1- F(\tau_{r}) \right)^{i_r} F(\tau_r)^{n- i_r}
\end{align*}
where the equality holds because both sides are the probability of having $i_r$ many values above $\tau_{r}$.

Given that 
\begin{align*}
    m \tau_r + S^+(\tau_r) \ge \mathrm{OPT}_m \text{ for all } r=1, \dots, k,
\end{align*}
and similar to the proof Theorem \ref{Thm:Comp:ratio}, we need to choose the sequence of thresholds to (i) maximize 
\begin{align*}
    \min\left\{ A\left(n-i_r, m-i_r, \frac{F(\tau_r)}{F(\tau_{r-1})}\right) , B\left(n-i_r, m-i_r, \frac{F(\tau_r)}{F(\tau_{r-1})}\right) \right\}
\end{align*}
and (ii) to make sure 
\begin{align*}
\binom{n}{i_r}    \left( 1- F(\tau_r)\right)^{i_r} F(\tau_r)^{n-i_r}
\end{align*}
is large enough. In the next lemma we establish the existence of thresholds that achieve both goals (i) and (ii). 

\begin{lemma}\label{Lem:polinomila:multipleitems}
For $x=1-\frac{m}{n}$, we have \begin{align*}
    \min\left\{ A(n, m, x), B(n, m, x) \right\} \ge 1- e^{-m} \frac{m^m}{m!}.
\end{align*}
Moreover, for any $\epsilon>0$, there exists $N(\epsilon)$ such that for $n \ge N(\epsilon)$:
\begin{align*}
    \binom{n}{i} \left( 1-x^r\right)^{i} x^{r(n-i)} \ge \frac{m^i r^i}{i!} \frac{e^{-mr}}{1+\epsilon} \text{ for all } i=0, \dots, m-1, r =1, \dots, k-1. 
\end{align*}
\end{lemma}

We defer the proof of this lemma to the appendix and continue with the proof of Theorem \ref{Thm:Comp:ratio:welfare:multiple} by using this lemma. 

We next lower bound the expression in \eqref{app:eq:pf:multipleitems:1}. First note that for all $r \in \{1, \dots, m\}$
\begin{align}\label{app:eq:pf:multipleitems:2}
    &m \tau_{r} A\left(n, m, \frac{F(\tau_{r})}{F(\tau_{r-1})}\right) + S^+(\tau_r) B\left(n, m, \frac{F(\tau_{r})}{F(\tau_{r-1})}\right) \nonumber \\
    & \ge  \min\left\{ A\left(n, m, \frac{F(\tau_{r})}{F(\tau_{r-1})}\right), B\left(n, m, \frac{F(\tau_{r})}{F(\tau_{r-1})}\right) \right\} \left( m \tau_r+ S^+(\tau_r) \right) \nonumber \\
    & = \min\left\{ A\left(n, m, \frac{F(\tau_{r})}{F(\tau_{r-1})}\right), B\left(n, m, \frac{F(\tau_{r})}{F(\tau_{r-1})}\right) \right\} \left(m \tau_r+\sum_{i=1}^n \mathbb{E}\left[ (V_i- \tau_1)^+\right] \right) \nonumber \\
    & \ge  \min\left\{ A\left(n, m, \frac{F(\tau_{r})}{F(\tau_{r-1})}\right), B\left(n, m, \frac{F(\tau_{r})}{F(\tau_{r-1})}\right) \right\} \mathrm{OPT}_m
\end{align}
Using Lemma \ref{Lem:polinomila:multipleitems} and \eqref{app:eq:pf:multipleitems:2}, the choice of
\begin{align*}
    F(\tau_r)= \left(1- \frac{m}{n}\right)^{r} \text{ for } r=1, \dots, k
\end{align*}
for $n \ge N(\epsilon)$ gives us approximation factor 
\begin{align*}
     \left(1- e^{-m} \frac{m^m}{m!} \right) + \sum_{r=1}^{k-1} \sum_{i=0}^{m-1} \frac{m^{i} r^{i}}{i!} \frac{e^{-mr}}{1+\epsilon} \left(1- e^{-(m-i)} \frac{(m-i)^{m-i}}{(m-i)!} \right).
\end{align*}
This completes the proof. \hfill\qedsymbol

We conclude this section by noting that Theorem \ref{Thm:Comp:ratio:welfare:multiple} covers the result of \cite{arnosti2021tight} that establishes $1-e^{-m} \frac{m^m}{m!}$ for a single threshold and a more involved  proof technique than our analysis.


\section{Conclusion}
\label{sec:conclusion}
Motivated by applications where the auctioneer is aiming to have less rounds of communications with buyers, we consider the descending price auction with a bounded number of price levels. As our main result, establish how well it can approximate the optimal revenue auction. In our problem formulation, an auctioneer with $m$ identical items posts $k$ prices and then multiple unit-demand buyers with i.i.d. values decide about their bids. To guide the analysis, we introduce a new variant of prophet inequality, called \emph{batched prophet inequality}, in which the decision-maker decides about $k$ (decreasing) thresholds and then sequentially collects rewards (up to $m$) that are above the thresholds by breaking ties uniformly at random. This variant of the classic prophet inequality is of independent interest, but we prove that the auctioneer's problem with bounded number of prices reduces to batched prophet inequality and then turn our attention to finding policies for the batched prophet inequality with optimal competitive ratio. For a single item, we establish the existence of a policy for the batched prophet inequality that achieves $1- 1/e^k$ of the optimal. Therefore, by increasing $k$, the revenue of a properly designed descending price auction with $k$ price levels converges exponentially fast to the optimal revenue. We then extend our analysis for the batched prophet inequality, and therefore descending price auction with $k$ bids, to a setting with multiple items. 



\bibliographystyle{plainnat}
\bibliography{refs}

\begin{thebibliography}{40}
\providecommand{\natexlab}[1]{#1}
\providecommand{\url}[1]{\texttt{#1}}
\expandafter\ifx\csname urlstyle\endcsname\relax
  \providecommand{\doi}[1]{doi: #1}\else
  \providecommand{\doi}{doi: \begingroup \urlstyle{rm}\Url}\fi

\bibitem[Abolhassani et~al.(2017)Abolhassani, Ehsani, Esfandiari, Hajiaghayi,
  Kleinberg, and Lucier]{abolhassani2017beating}
Melika Abolhassani, Soheil Ehsani, Hossein Esfandiari, MohammadTaghi
  Hajiaghayi, Robert Kleinberg, and Brendan Lucier.
\newblock Beating 1-1/e for ordered prophets.
\newblock In \emph{Proceedings of the 49th Annual ACM SIGACT Symposium on
  Theory of Computing}, pages 61--71, 2017.

\bibitem[Akbarpour and Li(2020)]{akbarpour2020credible}
Mohammad Akbarpour and Shengwu Li.
\newblock Credible auctions: A trilemma.
\newblock \emph{Econometrica}, 88\penalty0 (2):\penalty0 425--467, 2020.

\bibitem[Alaei et~al.(2019)Alaei, Hartline, Niazadeh, Pountourakis, and
  Yuan]{alaei2019optimal}
Saeed Alaei, Jason Hartline, Rad Niazadeh, Emmanouil Pountourakis, and Yang
  Yuan.
\newblock Optimal auctions vs. anonymous pricing.
\newblock \emph{Games and Economic Behavior}, 118:\penalty0 494--510, 2019.

\bibitem[Arnosti and Ma(2021)]{arnosti2021tight}
Nick Arnosti and Will Ma.
\newblock Tight guarantees for static threshold policies in the prophet
  secretary problem.
\newblock \emph{arXiv preprint arXiv:2108.12893}, 2021.

\bibitem[Azar et~al.(2014)Azar, Kleinberg, and Weinberg]{azar2014prophet}
Pablo~D Azar, Robert Kleinberg, and S~Matthew Weinberg.
\newblock Prophet inequalities with limited information.
\newblock In \emph{Proceedings of the twenty-fifth annual ACM-SIAM symposium on
  Discrete algorithms}, pages 1358--1377. SIAM, 2014.

\bibitem[Azar et~al.(2018)Azar, Chiplunkar, and Kaplan]{azar2018prophet}
Yossi Azar, Ashish Chiplunkar, and Haim Kaplan.
\newblock Prophet secretary: Surpassing the 1-1/e barrier.
\newblock In \emph{Proceedings of the 2018 ACM Conference on Economics and
  Computation}, pages 303--318, 2018.

\bibitem[Babaioff et~al.(2007)Babaioff, Immorlica, and
  Kleinberg]{babaioff2007matroids}
Moshe Babaioff, Nicole Immorlica, and Robert Kleinberg.
\newblock Matroids, secretary problems, and online mechanisms.
\newblock In \emph{Symposium on Discrete Algorithms (SODA'07)}, pages 434--443,
  2007.

\bibitem[Beyhaghi et~al.(2021)Beyhaghi, Golrezaei, Leme, P{\'a}l, and
  Sivan]{beyhaghi2018improved}
Hedyeh Beyhaghi, Negin Golrezaei, Renato~Paes Leme, Martin P{\'a}l, and
  Balasubramanian Sivan.
\newblock Improved revenue bounds for posted-price and second-price mechanisms.
\newblock \emph{Operations Research}, 69\penalty0 (6):\penalty0 1805--1822,
  2021.

\bibitem[Cai and Daskalakis(2011)]{cai2011extreme}
Yang Cai and Constantinos Daskalakis.
\newblock Extreme-value theorems for optimal multidimensional pricing.
\newblock In \emph{2011 IEEE 52nd Annual Symposium on Foundations of Computer
  Science}, pages 522--531. IEEE, 2011.

\bibitem[Chawla and Hartline(2013)]{chawla2013auctions}
Shuchi Chawla and Jason~D Hartline.
\newblock Auctions with unique equilibria.
\newblock In \emph{Proceedings of the fourteenth ACM conference on Electronic
  commerce}, pages 181--196, 2013.

\bibitem[Chawla et~al.(2007)Chawla, Hartline, and
  Kleinberg]{chawla2007algorithmic}
Shuchi Chawla, Jason~D Hartline, and Robert Kleinberg.
\newblock Algorithmic pricing via virtual valuations.
\newblock In \emph{Proceedings of the 8th ACM Conference on Electronic
  Commerce}, pages 243--251, 2007.

\bibitem[Chawla et~al.(2010)Chawla, Hartline, Malec, and
  Sivan]{chawla2009sequential}
Shuchi Chawla, Jason~D Hartline, David~L Malec, and Balasubramanian Sivan.
\newblock Multi-parameter mechanism design and sequential posted pricing.
\newblock In \emph{Proceedings of the forty-second ACM symposium on Theory of
  computing}, pages 311--320, 2010.

\bibitem[Chawla et~al.(2020)Chawla, Devanur, and Lykouris]{chawla2020static}
Shuchi Chawla, Nikhil Devanur, and Thodoris Lykouris.
\newblock Static pricing for multi-unit prophet inequalities.
\newblock \emph{arXiv preprint arXiv:2007.07990}, 2020.

\bibitem[Chwe(1989)]{chwe1989discrete}
Michael Suk-Young Chwe.
\newblock The discrete bid first auction.
\newblock \emph{Economics Letters}, 31\penalty0 (4):\penalty0 303--306, 1989.

\bibitem[Correa et~al.(2017)Correa, Foncea, Hoeksma, Oosterwijk, and
  Vredeveld]{correa2017posted}
Jos{\'e} Correa, Patricio Foncea, Ruben Hoeksma, Tim Oosterwijk, and Tjark
  Vredeveld.
\newblock Posted price mechanisms for a random stream of customers.
\newblock In \emph{Proceedings of the 2017 ACM Conference on Economics and
  Computation}, pages 169--186, 2017.

\bibitem[Correa et~al.(2019)Correa, Foncea, Hoeksma, Oosterwijk, and
  Vredeveld]{correa2019recent}
Jose Correa, Patricio Foncea, Ruben Hoeksma, Tim Oosterwijk, and Tjark
  Vredeveld.
\newblock Recent developments in prophet inequalities.
\newblock \emph{ACM SIGecom Exchanges}, 17\penalty0 (1):\penalty0 61--70, 2019.

\bibitem[Correa et~al.(2021)Correa, Saona, and Ziliotto]{correa2021prophet}
Jose Correa, Raimundo Saona, and Bruno Ziliotto.
\newblock Prophet secretary through blind strategies.
\newblock \emph{Mathematical Programming}, 190\penalty0 (1):\penalty0 483--521,
  2021.

\bibitem[Dutting et~al.(2020)Dutting, Feldman, Kesselheim, and
  Lucier]{dutting2020prophet}
Paul Dutting, Michal Feldman, Thomas Kesselheim, and Brendan Lucier.
\newblock Prophet inequalities made easy: Stochastic optimization by pricing
  nonstochastic inputs.
\newblock \emph{SIAM Journal on Computing}, 49\penalty0 (3):\penalty0 540--582,
  2020.

\bibitem[Ehsani et~al.(2018)Ehsani, Hajiaghayi, Kesselheim, and
  Singla]{ehsani2018prophet}
Soheil Ehsani, MohammadTaghi Hajiaghayi, Thomas Kesselheim, and Sahil Singla.
\newblock Prophet secretary for combinatorial auctions and matroids.
\newblock In \emph{Proceedings of the twenty-ninth annual acm-siam symposium on
  discrete algorithms}, pages 700--714. SIAM, 2018.

\bibitem[Esfandiari et~al.(2017)Esfandiari, Hajiaghayi, Liaghat, and
  Monemizadeh]{esfandiari2017prophet}
Hossein Esfandiari, MohammadTaghi Hajiaghayi, Vahid Liaghat, and Morteza
  Monemizadeh.
\newblock Prophet secretary.
\newblock \emph{SIAM Journal on Discrete Mathematics}, 31\penalty0
  (3):\penalty0 1685--1701, 2017.

\bibitem[Feldman et~al.(2016)Feldman, Svensson, and
  Zenklusen]{feldman2016online}
Moran Feldman, Ola Svensson, and Rico Zenklusen.
\newblock Online contention resolution schemes.
\newblock In \emph{Proceedings of the twenty-seventh annual ACM-SIAM symposium
  on Discrete algorithms}, pages 1014--1033. SIAM, 2016.

\bibitem[Haghpanah and Hartline(2015)]{haghpanah2015reverse}
Nima Haghpanah and Jason Hartline.
\newblock Reverse mechanism design.
\newblock In \emph{Proceedings of the Sixteenth ACM Conference on Economics and
  Computation}, pages 757--758, 2015.

\bibitem[Hartline(2012)]{hartline2012approximation}
Jason~D Hartline.
\newblock Approximation in mechanism design.
\newblock \emph{American Economic Review}, 102\penalty0 (3):\penalty0 330--36,
  2012.

\bibitem[Hartline(2013)]{hartline2013mechanism}
Jason~D Hartline.
\newblock Mechanism design and approximation.
\newblock \emph{Book draft. October}, 122:\penalty0 1, 2013.

\bibitem[Hartline and Roughgarden(2009)]{hartline2009simple}
Jason~D Hartline and Tim Roughgarden.
\newblock Simple versus optimal mechanisms.
\newblock In \emph{Proceedings of the 10th ACM conference on Electronic
  commerce}, pages 225--234, 2009.

\bibitem[Hill and Kertz(1982)]{hill1982comparisons}
Theodore~P Hill and Robert~P Kertz.
\newblock Comparisons of stop rule and supremum expectations of iid random
  variables.
\newblock \emph{The Annals of Probability}, pages 336--345, 1982.

\bibitem[H{\"o}rner and Samuelson(2011)]{horner2011managing}
Johannes H{\"o}rner and Larry Samuelson.
\newblock Managing strategic buyers.
\newblock \emph{Journal of Political Economy}, 119\penalty0 (3):\penalty0
  379--425, 2011.

\bibitem[Jiang et~al.(2022)Jiang, Ma, and Zhang]{jiang2022tight}
Jiashuo Jiang, Will Ma, and Jiawei Zhang.
\newblock Tight guarantees for multi-unit prophet inequalities and online
  stochastic knapsack.
\newblock In \emph{Proceedings of the 2022 Annual ACM-SIAM Symposium on
  Discrete Algorithms (SODA)}, pages 1221--1246. SIAM, 2022.

\bibitem[Jin et~al.(2019)Jin, Lu, Qi, Tang, and Xiao]{jin2019tight}
Yaonan Jin, Pinyan Lu, Qi~Qi, Zhihao~Gavin Tang, and Tao Xiao.
\newblock Tight approximation ratio of anonymous pricing.
\newblock In \emph{Proceedings of the 51st Annual ACM SIGACT Symposium on
  Theory of Computing}, pages 674--685, 2019.

\bibitem[Kleinberg and Weinberg(2012)]{kleinberg2012matroid}
Robert Kleinberg and Seth~Matthew Weinberg.
\newblock Matroid prophet inequalities.
\newblock In \emph{Proceedings of the forty-fourth annual ACM symposium on
  Theory of computing}, pages 123--136, 2012.

\bibitem[Krengel and Sucheston(1978)]{krengel1978semiamarts}
Ulrich Krengel and Louis Sucheston.
\newblock On semiamarts, amarts, and processes with finite value.
\newblock \emph{Probability on Banach spaces}, 4:\penalty0 197--266, 1978.

\bibitem[Lee and Singla(2018)]{lee2018optimal}
Euiwoong Lee and Sahil Singla.
\newblock Optimal online contention resolution schemes via ex-ante prophet
  inequalities.
\newblock In \emph{26th European Symposium on Algorithms, ESA 2018}. Schloss
  Dagstuhl-Leibniz-Zentrum fur Informatik GmbH, Dagstuhl Publishing, 2018.

\bibitem[Liu et~al.(2021)Liu, Leme, P{\'a}l, Schneider, and
  Sivan]{liu2020variable}
Allen Liu, Renato~Paes Leme, Martin P{\'a}l, Jon Schneider, and Balasubramanian
  Sivan.
\newblock Variable decomposition for prophet inequalities and optimal ordering.
\newblock In \emph{Proceedings of the 22nd ACM Conference on Economics and
  Computation}, pages 692--692, 2021.

\bibitem[Lucier(2017)]{lucier2017economic}
Brendan Lucier.
\newblock An economic view of prophet inequalities.
\newblock \emph{ACM SIGecom Exchanges}, 16\penalty0 (1):\penalty0 24--47, 2017.

\bibitem[Milgrom(2004)]{milgrom2004putting}
Paul Milgrom.
\newblock \emph{Putting Auction Theory to Work}.
\newblock Cambridge University Press, 2004.

\bibitem[Myerson(1981)]{myerson1981optimal}
Roger~B Myerson.
\newblock Optimal auction design.
\newblock \emph{Mathematics of operations research}, 6\penalty0 (1):\penalty0
  58--73, 1981.

\bibitem[Nguyen and Sandholm(2014)]{nguyen2014optimizing}
Tri-Dung Nguyen and Tuomas Sandholm.
\newblock Optimizing prices in descending clock auctions.
\newblock In \emph{Proceedings of the fifteenth ACM conference on Economics and
  computation}, pages 93--110, 2014.

\bibitem[Samuel-Cahn(1984)]{samuel1984comparison}
Ester Samuel-Cahn.
\newblock Comparison of threshold stop rules and maximum for independent
  nonnegative random variables.
\newblock \emph{the Annals of Probability}, pages 1213--1216, 1984.

\bibitem[Vickrey(1961)]{vickrey1961counterspeculation}
William Vickrey.
\newblock Counterspeculation, auctions, and competitive sealed tenders.
\newblock \emph{The Journal of finance}, 16\penalty0 (1):\penalty0 8--37, 1961.

\bibitem[Yan(2011)]{yan2011mechanism}
Qiqi Yan.
\newblock Mechanism design via correlation gap.
\newblock In \emph{Proceedings of the twenty-second annual ACM-SIAM symposium
  on Discrete Algorithms}, pages 710--719. SIAM, 2011.

\end{thebibliography}
\newpage

\appendix
\section{Omitted Proofs}
\subsection*{Proof of \Cref{prop:BNE}}
\label{apx:proof-prop-BNE}
To show part (i) of \Cref{prop:BNE}, it is enough to show that the mapping from any buyer's value to its equilibrium bid, denoted by $\eqbid(v)$, is monotone non-decreasing in $v$. This implies that $\eqbid(v)$ is a non-decreasing step function, and that the equilibrium can be identified by a sequence of weakly decreasing thresholds $\tau_1\geq\tau_2\geq\ldots\tau_k$, as described in the statement of part (i) of \Cref{prop:BNE}. Note that we only show $\eqbid(v)$ is a step function with $k'\leq k$ distinct steps, with step values being a strictly decreasing sub-sequence of $p_1>p_2>\ldots>p_k$. In principle, our price sequence might not be minimal, meaning that $k'<k$.


Fix the bidding strategies of all buyers in $\buyerset$ except $i$. Consider values $v_1<v_2$ for buyer $i$. Let $b_1=\eqbid(v_1)$ and $b_2=\eqbid(v_2)$. Moreover, let $x_1$ and $x_2$ be the allocation probabilities for buyer $i$ given bids $b_1$ and $b_2$, respectively. First suppose buyer $i$'s value is realized to be $v_1$. Because $b_1$ is her best-response bid under value $v_1$, we have:
$$
x_1(v_1-b_1)\geq x_2(v_1-b_2)
$$
Now suppose buyer $i$'s value is realized to be $v_2$. Because $b_2$ is her best-response bid under value $v_2$, we have:
$$
x_2(v_2-b_2)\geq x_1(v_2-b_1).
$$
Summing up the above inequalities and rearranging the terms, we have:
$$
(x_1-x_2)(v_1-v_2)\geq 0
$$
Therefore, as $v_1<v_2$, we should have $x_1\leq x_2$. Note that in $k$-\dca{} the allocation probability of buyer $i$ as a function of her submitted bid is increasing. Therefore, $b_1\leq b_2$, as desired.

To show part (ii) of \Cref{prop:BNE}, we develop the indifference condition for buyers. To be more formal, suppose all other buyers except buyer 1 play with the BNE strategy $\eqbid$ in part (i). Now suppose buyer 1's value is $v=\tau_j+\epsilon$ for small enough $\epsilon>0$. Then the expected utility of such a buyer when selecting price $p_j$ should be no more than when selecting price $p_{j+1}$. Now suppose buyer 1's value is $v=\tau_j-\epsilon$ for small enough $\epsilon>0$. Then the expected utility of such a buyer when selecting price $p_{j+1}$ should be no more than when selecting price $p_{j}$. Taking the limit as $\epsilon\rightarrow 0$ indicates that the expected utility of a buyer with value $\tau_j$ should be the same under bidding either $p_j$ or $p_{j+1}$. 

Now suppose round $j$ with price $p_j$ has arrived. If a buyer with value $v$ accepts this price, her expected utility will be 
\begin{align*}
     & \sum_{\ell=0}^{n-1} \frac{1}{\ell+1} \binom{n-1}{\ell} \left( 1- \frac{\buyerCDF(\tau_j)}{\buyerCDF(\tau_{j-1})}\right)^{\ell}  \left(\frac{\buyerCDF(\tau_j)}{\buyerCDF(\tau_{j-1})}\right)^{n-1-\ell} (v - p_j)   \\
     &= \frac{1}{\left( 1- \frac{\buyerCDF(\tau_j)}{\buyerCDF(\tau_{j-1})}\right)}\sum_{\ell=0}^{n-1} \frac{1}{n} \binom{n}{\ell+1} \left( 1- \frac{\buyerCDF(\tau_j)}{\buyerCDF(\tau_{j-1})}\right)^{\ell+1}  \left(\frac{\buyerCDF(\tau_j)}{\buyerCDF(\tau_{j-1})}\right)^{n-1-\ell} (v - p_j)   \\
     &= \frac{1}{\left( 1- \frac{\buyerCDF(\tau_j)}{\buyerCDF(\tau_{j-1})}\right)}\sum_{\ell=1}^{n} \frac{1}{n} \binom{n}{\ell} \left( 1- \frac{\buyerCDF(\tau_j)}{\buyerCDF(\tau_{j-1})}\right)^{\ell}  \left(\frac{\buyerCDF(\tau_j)}{\buyerCDF(\tau_{j-1})}\right)^{n-\ell} (v - p_j)   \\
     & = \frac{1- \left(\frac{\buyerCDF(\tau_j)}{\buyerCDF(\tau_{j-1})} \right)^n}{1- \frac{\buyerCDF(\tau_j)}{\buyerCDF(\tau_{j-1})}} \frac{v - p_j}{n}.
\end{align*}
If she waits and accepts the price $p_{j+1}$ in round $j+1$, her utility will become 
\begin{align*}
   & \left(\frac{\buyerCDF(\tau_{j})}{\buyerCDF(\tau_{j-1})} \right)^{n-1}  \sum_{\ell=0}^{n-1} \frac{1}{\ell+1} \binom{n-1}{\ell} \left( 1- \frac{\buyerCDF(\tau_{j+1})}{\buyerCDF(\tau_{j})}\right)^{\ell}  \left(\frac{\buyerCDF(\tau_{j+1})}{\buyerCDF(\tau_{j})}\right)^{n-1-\ell} (v - p_{j+1})   \\
   & = \left(\frac{\buyerCDF(\tau_{j})}{\buyerCDF(\tau_{j-1})} \right)^{n-1} \frac{1- \left(\frac{\buyerCDF(\tau_{j+1})}{\buyerCDF(\tau_{j})} \right)^n}{1- \frac{\buyerCDF(\tau_{j+1})}{\buyerCDF(\tau_{j})}} \frac{v - p_{j+1}}{n}.
\end{align*}
The indifference condition then implies that the above two utilities are equal for $v= \tau_{j}$, which leads to the equations in the statement of the proposition. \hfill\qedsymbol
\subsection*{Proof of \Cref{Pro:PriceToThresholds:Surjective-new}}
\label{apx:proof-of-inverse-map}
Suppose thresholds $\tau_1>\ldots>\tau_k>0$ are given. Viewing prices as variables, we show there exists a unique solution $\pbf=(p_1,\ldots,p_k)$ to the system of linear equations defined by our indifference conditions, i.e., $p_k=\tau_k$  and \cref{Eq:Pro:PricesToThresholds-new} for $j\in[k-1]$, so that $p_1>\ldots>p_k>0$. Rearranging the terms in \cref{Eq:Pro:PricesToThresholds-new}, we have:
\begin{align}\label{Eq:Pro:PricesToThresholds-rearrange}
  \frac{\left(\buyerCDF(\tau_{j-1})\right)^{n}- \left(\buyerCDF(\tau_{j})\right)^{n}}{\buyerCDF(\tau_{j-1})-\buyerCDF(\tau_j)} \frac{\tau_{j} - p_j}{n}=\frac{\left(\buyerCDF(\tau_{j})\right)^{n}- \left(\buyerCDF(\tau_{j+1})\right)^{n}}{\buyerCDF(\tau_{j})-\buyerCDF(\tau_{j+1})}  \frac{\tau_{j} - p_{j+1}}{n},~
\end{align}
or equivalently
\begin{align}\label{Eq:Pro:PricesToThresholds-rearrange-2}
\left(\sum_{i=0}^{n-1} \left(\buyerCDF(\tau_{j-1})\right)^{i}\left(\buyerCDF({\tau_{j})}\right)^{n-1-i}\right)(\tau_j-p_j)=\left(\sum_{i=0}^{n-1} \left(\buyerCDF(\tau_{j})\right)^{i}\left(\buyerCDF({\tau_{j+1})}\right)^{n-1-i}\right)(\tau_j-p_{j+1}).
\end{align}
Starting from $p_k=\tau_k$, we can recursively construct a sequence of prices that satisfy \cref{Eq:Pro:PricesToThresholds-rearrange-2} for $j\in[k-1]$. As $\buyerCDF(\tau_{j-1})>\buyerCDF(\tau_{j})$ and $\buyerCDF(\tau_{j})>\buyerCDF(\tau_{j+1})$, we have $\tau_j-p_j<\tau_j-p_{j+1}$, and hence $p_{j+1}<p_{j}$. As a result, $\pbf$ is a sequence of strictly decreasing prices whose corresponding symmetric BNE is identified by thresholds $\boldsymbol\tau$, as desired. \hfill\qedsymbol

\subsection*{Proof of Lemma \ref{Lem:inequality}}
\label{apx:lemma:balance}
We can write 
\begin{align*}
    P_n(x)= \sum_{i=0}^{n} \frac{1}{i+1} \binom{n-1}{i} x^{n-1-i} (1-x)^{i} = \frac{1}{n (1- x)} (1-x^n),
\end{align*}
which is increasing in $x$. The function $1-x^n$ is decreasing in $x$. Making these two equal results in 
\begin{align*}
    x= 1- \frac{1}{n}.
\end{align*}
We now evaluate $1- x^n$ (and therefore $P_n(x)$) at this point:
\begin{align*}
    1- \left( 1- \frac{1}{n} \right)^n \ge 1- \frac{1}{e}.
\end{align*}
This establishes for $x=1-\frac{1}{n}$
\begin{align*}
    \min\{1-x^n, P_n(x) \} \ge 1- \frac{1}{e}.
\end{align*}
However, $x^n$ is not necessarily large. We next show how we can choose another $x$ that satisfies both criteria.

For $x=\left(1/e \right)^{1/n}$, we have 
\begin{align*}
    x^n = \frac{1}{e}, \quad 1- x^n= 1- \frac{1}{e},
\end{align*}
and 
\begin{align*}
     P_n(x)= \left(1- \frac{1}{e} \right) \frac{1}{n (1- \left(1/e \right)^{1/n})} \overset{(a)}{\ge} 1- \frac{1}{e},
\end{align*}
where (a) follows from 
\begin{align*}
    \left( 1- \frac{1}{n} \right)^n \le \frac{1}{e}.
\end{align*}
This completes the proof. \qed

\subsection*{Proof of Theorem \ref{Thm:Comp:ratio}}
\label{apx:proof-thm-comp-ratio}
We let $\algT(\tau_1, \dots, \tau_k)$ denote the expected reward of the sequence of thresholds $\tau_1, \dots, \tau_k$, we can write  
\begin{align}\label{eq:pf:bound}
    &\mathbb{E}[\algT(\tau_1, \dots, \tau_k)] \overset{(a)}{=} \left(1- F(\tau_1)^n \right) \tau_1 \nonumber \\
    &+ \sum_{i=1}^n \mathbb{E}\left[(V_i -\tau_1)^+ \right] \sum_{j_1=0}^{n-1} \binom{n-1}{j_1} \left(1- F(\tau_1) \right)^{j} F(\tau_1)^{n-1-j_1} \frac{1}{1+j_1} \nonumber \\
    & + F(\tau_1)^n \Biggl(\left(1- \left(\frac{F(\tau_2)}{F(\tau_1)}\right)^n \right)  \tau_2  \nonumber \\
    & + \sum_{i=1}^n \mathbb{E}\left[(V_i -\tau_1)^+ \right] \sum_{j_2=0}^{n-1} \binom{n-1}{j_2} \left(1- \left(\frac{F(\tau_2)}{F(\tau_1)}\right)^{j} \right) \left(\frac{F(\tau_2)}{F(\tau_1)}\right)^{n-1-j_2} \frac{1}{1+j_2}  + \cdots \Biggr) \nonumber \\
    &\overset{(b)}{=} 
    \left(1- F(\tau_1)^n \right) \tau_1 + P_n\left(F(\tau_1)\right) \sum_{i=1}^n \mathbb{E}\left[(V_i -\tau_1)^+ \right] \nonumber \\
    & + F(\tau_1)^n  \Bigl( \left(1- \left(\frac{F(\tau_2)}{F(\tau_1)}\right)^n \right) \tau_2 + P_n\left(\frac{F(\tau_2)}{F(\tau_1)}\right) \sum_{i=1}^n \mathbb{E}\left[(V_i -\tau_2)^+ \right] + \cdots \Bigr).
\end{align}
where (a) follows from the fact that when $j$ many rewards are above $\tau_1$ the algorithm captures reward $\mathbb{E}[(V_i - \tau_1)^+]$ with probability $\frac{1}{1+j}$ and (b)  follows from the definition of polynomial $P_n(\cdot)$. Using Lemma \ref{Lem:inequality}, for 
\begin{align*}
F(\tau_1)=\left(\frac{1}{e}\right)^{\frac{1}{n}}
\end{align*}
we have 
\begin{align*}
    \left(1- F(\tau_1)^n \right) \tau_1 + P_n\left(F(\tau_1)\right) \sum_{i=1}^n \mathbb{E}\left[(V_i -\tau_1)^+ \right]  \ge \left(1- \frac{1}{e}\right)\left(\tau_1+\sum_{i=1}^n \mathbb{E}[(V_i-\tau_1)^+]\right)\geq \left(1- \frac{1}{e}\right)\mathrm{OPT}
\end{align*}
and also 
\begin{align*}
    F(\tau_1)^n = \frac{1}{e}.
\end{align*}
By using Lemma \ref{Lem:inequality} multiple times, and using the fact that $\tau+\sum_{i=1}^n \mathbb{E}[(V_i-\tau)^+]\geq \maxV$ for any $\tau$, 
we can further bound \eqref{eq:pf:bound} as 
\begin{align*}
    & \left(1- \frac{1}{e} \right) \mathrm{OPT} + \frac{1}{e} \left( \left(1- \frac{1}{e} \right) \mathrm{OPT}+ \dots \right)  \\
    & \ge \left(1- \frac{1}{e} \right) \mathrm{OPT} + \frac{1}{e} \left( \left(1- \frac{1}{e} \right) \mathrm{OPT} + \frac{1}{e} \cdots   \right)\\
    & = \mathrm{OPT} \left(1- \frac{1}{e^k} \right),
\end{align*}
which completes the proof. \qed

\subsection*{Proof of \Cref{Thm:Comp:ratio:revenue}}
\label{apx:proof-thm-revenue}
Given buyer values $v_1,\ldots,v_n\sim G$ in the $k$-\dca{} problem, consider an instance of the batched prophet inequality problem with i.i.d. rewards $V_i=\virtualV(v_i)$, where $V_i\sim F$ and $F(x)=G\left(\virtualV^{-1}(x)\right)$. Define the randomized set $\hat{\mathcal{N}}=\{i\in[n]:v_i\geq \virtualV^{-1}(0)\}$. Buyers in $\hat{\mathcal{N}}$ are the only buyers whose virtual value is non-negative. Therefore $$\maxV=\mathbb{E}\left[\max_{i\in\hat{\buyerset}}V_i\right].$$
Now fix a particular realization  $\hat{\buyerset}=S$ for some $S\subseteq [n]$. Note that conditioned on the event $\hat{\buyerset}=S$, as $V_i$'s are i.i.d., any $V_i$ for $i\in S$ is drawn from the same conditional CDF 
$$\tilde{F}(x)=\mathbb{P}\left[V_i\leq x \mid V_i\geq 0\right]=\left(F(x)-F(0)\right)/(1-F(0))=\left(G\left(\virtualV^{-1}(x)\right)-G(\rho)\right)/(1-G(\rho))~,$$
where $\rho=\virtualV^{-1}(0)$. We now invoke \Cref{Thm:Comp:ratio} with rewards being $\lvert S\rvert$ i.i.d. non-negative random variables drawn from distribution $\tilde{F}(x)$ (with support $[0,+\infty))$, in order to obtain $k$ thresholds $\tau_1>\ldots>\tau_k>0$. Based on the construction of these thresholds, we can write
$$
\tilde{F}(\tau_j)=\frac{1}{e^{j/n}}~~,~~\textrm{for}~~j=1,\ldots,k~,
$$
and therefore we have:
\begin{equation}
\label{eq:revenue-thresh}
G\left(\virtualV^{-1}(\tau_j)\right)=\frac{1-G(\rho)}{e^{j/n}}+G(\rho)~~,~~\textrm{for}~~j=1,\ldots,k.
\end{equation}
Note that the choice of these thresholds \emph{does not} depend on the exact realization $S$. Now, by
applying the competitive ratio lower-bound of \Cref{Thm:Comp:ratio} for this instance, we have:
$$
\mathbb{E}\left[\algT(\tau_1,\ldots,\tau_k) \mid \hat\buyerset=S\right]\geq \left(1-\frac{1}{e^k}\right)
\mathbb{E}\left[\max_{i\in S}V_i \mid \hat\buyerset=S\right]~.$$
By taking expectation over $\hat{\buyerset}$, we have 
$$
\mathbb{E}\left[\algT(\tau_1,\ldots,\tau_k) \right]\geq \left(1-\frac{1}{e^k}\right)
\mathbb{E}\left[\max_{i\in \hat{\buyerset}}V_i \right]=\left(1-\frac{1}{e^k}\right)
\maxV~.
$$

Now, applying the reduction in \Cref{sec:reduction}, we know if (i) we set $\hat{\tau}_j=\virtualV^{-1}(\tau_j)$ and  use $\{\hat{\tau}_j\}_{j\in[k]}$ as the equilibrium thresholds of a $k$-\dca{} against the original buyers, and then (ii) we use \Cref{Pro:PriceToThresholds:Surjective-new} and \Cref{Eq:Pro:PricesToThresholds-new} to obtain prices $\pbf$ supporting these equilibrium thresholds, then:
$$
\mathbb{E}\left[\rev{\vbf}{\pbf}\right]=\mathbb{E}\left[\algT(\tau_1,\ldots,\tau_k)\right]~.
$$
Moreover, the expected revenue of Myerson's optimal mechanism is equal to:
$$
\mathbb{E}\left[\opt{\vbf}\right]=\maxV~,
$$
and hence using prices $\pbf$ in $k$-\dca{} will result in the desired approximation ratio. 

Now, from \cref{eq:revenue-thresh} and the fact that $\hat{\tau}_j=\virtualV^{-1}(\tau_j)$ we have:
\begin{equation}
\label{eq:pf:thresholds:unifomr}
G\left(\hat\tau_j\right)=\left(1-G(\rho)\right)\alpha^j+G(\rho)~~,~~\textrm{for}~~j=1,\ldots,k~, \textrm{where}~\alpha=\frac{1}{e^{1/n}}~.
\end{equation}
Using \Cref{Pro:PriceToThresholds:Surjective-new} and \Cref{Eq:Pro:PricesToThresholds-new} for the sequence of thresholds given in \eqref{eq:pf:thresholds:unifomr}, the sequence of prices must satisfy
\begin{equation*}
\frac{\left(\buyerCDF(\hat\tau_{j-1})\right)^{n}- \left(\buyerCDF(\tau_{j})\right)^{n}}{\buyerCDF(\hat\tau_{j-1})-\buyerCDF(\hat\tau_j)} \left(\hat\tau_{j} - p_j\right)=\frac{\left(\buyerCDF(\hat\tau_{j})\right)^{n}- \left(\buyerCDF(\hat\tau_{j+1})\right)^{n}}{\buyerCDF(\hat\tau_{j})-\buyerCDF(\hat\tau_{j+1})}  \left(\hat\tau_{j} - p_{j+1}\right),~
\end{equation*}
with the initialization $p_k=G^{-1}\left(\left(1-G(\rho)\right)\alpha^k+G(\rho)\right)$. Rearranging the terms gives us the update equation of Algorithm~\ref{alg:prices-revenue}, which completes the proof.\qed
\subsection*{Proof of Proposition \ref{Pro:PricesToThresholds:multipleitems}}
The proof of this proposition is similar to that of Proposition \ref{prop:BNE}. We next develop the indifference condition for buyers. If a buyer with value $v$ accepts this price, her expected utility will be 
\begin{align*}
     & \sum_{r=0}^m \sum_{\ell=0}^{n-1-r} \binom{n-1}{r} \left(1-F(\tau_{j-1}) \right)^r F(\tau_{j-1})^{n-1-r} \min\left\{\frac{m-r}{\ell+1},1\right\} \binom{n-1-r}{\ell} \\
     & \left( 1- \frac{F(\tau_j)}{F(\tau_{j-1})}\right)^{\ell}  \left(\frac{F(\tau_j)}{F(\tau_{j-1})}\right)^{n-1-r-\ell} (v - p_j)   
\end{align*}
If she accepts the price $p_{j+1}$ in round $j+1$, her expected utility  becomes
\begin{align*}
     & \sum_{r=0}^m \sum_{\ell=0}^{n-1-r} \binom{n-1}{r} \left(1-F(\tau_{j}) \right)^r F(\tau_{j})^{n-1-r} \min\left\{\frac{m-r}{\ell+1},1\right\} \binom{n-1-r}{\ell} \\
     & \left( 1- \frac{F(\tau_{j+1})}{F(\tau_{j})}\right)^{\ell}  \left(\frac{F(\tau_{j+1})}{F(\tau_{j})}\right)^{n-1-r-\ell} (v - p_{j+1})   
\end{align*}
The indifference condition then implies that the above two utilities are equal for $v= \tau_{j}$, which leads to the equations in the statement of the proposition. \hfill\qedsymbol


\subsection*{Proof of Lemma \ref{Lem:polinomila:multipleitems}}
We first establish that 
\begin{align}\label{app:eq:Lem:polinomila:multipleitems}
    B(n, m, x)= \frac{m A(n, m, x)}{n(1-x)}.
\end{align}
To see this notice that 
\begin{align*}
    B(n, m, x) = & \sum_{i=0}^{n-1} x^{n-1-i} (1-x)^i \binom{n-1}{i} \min\left\{1, \frac{m}{i+1}\right\} \\
     \overset{(a)}{=} & \sum_{i=0}^{n-1} x^{n-1-i} (1-x)^i \binom{n}{i+1} \frac{1}{n} \min\left\{i+1, m\right\} \\
     \overset{(b)}{=} & \sum_{i=1}^{n} x^{n-i} (1-x)^{i-1} \binom{n}{i} \frac{1}{n} \min\left\{i, m\right\} = \frac{m A(n, m, x)}{n(1-x)},
\end{align*}
where (a) follows from 
\begin{align*}
    \binom{n-1}{i} = \binom{n}{i+1} \frac{i+1}{n}
\end{align*}
and (b) follows from a change of variable from $i$ to $i+1$.

For $x=1- \frac{m}{n}$ and $i \le m-1$ we have 
\begin{align*}
    \binom{n}{i} \left( 1-x^r\right)^{i} x^{r(n-i)}= & \binom{n}{i} \left(\frac{1-\left(1- \frac{m}{n} \right)^{r}}{\left(1- \frac{m}{n} \right)^{r}} \right)^{i} \left(1- \frac{m}{n} \right)^{rn} \\
    & \overset{(a)}{\ge}  \left(1- \frac{m}{n} \right)^{rn}\frac{1}{i!}  \left((n-i+1) \frac{1-\left(1- \frac{m}{n} \right)^{r}}{\left(1- \frac{m}{n} \right)^{r}} \right)^{i} \\
    & \overset{(b)}{\ge}  \left(1- \frac{m}{n} \right)^{rn}\frac{1}{i!}  \left((n-m) \frac{1-\left(1- \frac{m}{n} \right)^{r}}{\left(1- \frac{m}{n} \right)^{r}} \right)^{i} \\
    & \overset{(c)}{\ge}  \left(1- \frac{m}{n} \right)^{rn} \frac{1}{i!}  \left(mr \right)^{i} \\
    & \overset{(d)}{\ge}  \frac{e^{-mr}}{1+\epsilon} \frac{1}{i!}  \left(mr \right)^{i}  
\end{align*}
where (a) follows from $\frac{n!}{(n-i)!} \ge (n-i+1)^i$, (b) follows from $i \le m-1$, (c) follows from 
\begin{align*}
    \left(1- \frac{m}{n} \right)^{r} \le e^{-\frac{rm}{n}}
\end{align*}
and
\begin{align*}
    e^{rm} \ge e^{\frac{rmn}{n-m}} \ge  \left(1+ \frac{rm}{n-m} \right)^n 
\end{align*}
which in turn implies 
\begin{align*}
    \frac{1-\left(\frac{1}{e} \right)^{\frac{rm}{n}}}{\left(\frac{1}{e} \right)^{\frac{rm}{n}}}= e^{\frac{rm}{n}} -1 \ge \frac{mr}{n-m}.
\end{align*}
Finally, (d) follows from the fact that the limit of $\left(1- \frac{m}{n} \right)^{rn}$ as $n \to \infty$ is $e^{-mr}$ and $n$ is large enough. We need to pick $N(\epsilon)$ large enough such that (d) holds for all $i=0, \dots, m-1$ and $r=1, \dots, k$. This is possible because we can take $N(\epsilon)$ the maximum of finitely many $N(\epsilon, r, k)$.

We next show that for $x=1- \frac{m}{n}$, we have 
\begin{align*}
    B(n, m, x) \ge 1- e^{-m} \frac{m^m}{m!} \text{ and } A(n, m, x) \ge 1- e^{-m} \frac{m^m}{m!}
\end{align*}
Using \eqref{app:eq:Lem:polinomila:multipleitems}, it suffices to show
\begin{align*}
    A(n, m, x) \ge 1- e^{-m} \frac{m^m}{m!}.
\end{align*}
Letting $x=1-m/n$, we can write 
\begin{align*}
     A\left(n, m, \left(1- \frac{m}{n} \right)\right) =& \sum_{i=1}^m x^{n-i} (1- x)^i \binom{n}{i} \frac{i}{m} + \sum_{i=m+1}^n x^{n-i} (1- x)^i \binom{n}{i} \\
     & = 1- \sum_{i=0}^m x^{n-i} (1-x)^i \binom{n}{i} \left( 1- \frac{i}{m}\right) \\
     & \overset{(a)}{\ge} 1- \sum_{i=0}^m  e^{-m} \frac{m^i}{i!} \left( 1- \frac{i}{m}\right) \\
     & \overset{(b)}{=} 1- e^{-m} \frac{m^m}{m!}.
\end{align*}
where (a) follows from \citet[Lemma 4.2]{yan2011mechanism} and the fact that the right-hand side is the limit as $n \to \infty$ and (b) holds because we have a telescopic summation. This completes the proof. \hfill\qedsymbol

\end{document}